\documentclass[twocolumn, floatfix,secnumarabic, amssymb, nobibnotes, aps, prd]{revtex4-2}

\usepackage{mathrsfs}

\usepackage{amssymb}

\usepackage{graphicx, bm}
\begin{document}

\title{Derivation of Hamiltonians from time propagations using Born machines.}%

\author{Hikaru Wakaura}%
\email[Quantscape: ]{
hikaruwakaura@gmail.com}
\affiliation{QuantScape Inc. QuantScape Inc., 4-11-18, Manshon-Shimizudai, Meguro, Tokyo, 153-0064, Japan}

\author{Andriyan B. Suksmono}

\affiliation{ Institut Teknologi Bandung, Jl. Ganesha No.10, Bandung, Jawa Barat, Indonesia}
\email[Bandung Institute of Technology: ]{suksmono@stei.itb.ac.id}

\begin{abstract}

Recently, there are more promising qubit technology--such as Majorana fermions, Rydberg atoms, and Silicon quantum dot, have yet to be developed for realizing a quantum computer than Superconductivity and Ion-trap into the world.

The simulation of the quantum hardwares of these qubits can only be done numerically. However, a classical numerical simulation is limited concerning available resources. The method for simulation of quantum hardware by quantum hardware may be necessary. 
In this paper, we propose a novel method for optimizing time propagation from initial states to aimed given states of systems by the Born machine.
We call this method the Hamiltonian Engineering Born Machine (HEBM). We calculated the optimal Hamiltonians for propagation to Bars and Stripes distribution, Gaussian distribution, and Gibbs state for $H = - \sum_{j = 0}Z_j Z_{j + 1}$ and revealed that they can be realized rapidly and accurately.

\end{abstract}
\maketitle
\tableofcontents

\section{Introduction}\label{1}

In 1982, Richard P. Feynman proposed the theory of quantum computers at first\cite{feynman_simulating_1982}. Since then, many quantum algorithms that take advantage of quantum computing have been brought into the world, such as Grover's\cite{2003quant.ph..1079L} and Shor's\cite{365700} algorithm. Although, quantum systems that can be used for qubits and quantum computers do not have enough properties for practical use. So, the research of quantum algorithms had not advanced rapidly until 2015, when IBM released available quantum computers on cloud platforms. Superconducting and Ion-Trap quantum computers are major practical available hardwares. However, many candidates of qubits can realize the more small-sized quantum computers and a large number of qubits such as Majorana fermions\cite{Mezzacapo_2013}, Nitro Vacancy of diamond lattice\cite{doi:10.1126/science.1139831}, and Rydberg atoms\cite{2020arXiv201210614W}. However, their research is still ongoing in the area of quantum computers. A method that can simulate and optimize the time propagation of quantum systems is needed.
On the other hand, a lot of quantum algorithms have been developed since 2015. Dr. Aran-Aspuru Guzik proposed the foundation of Variational Quantum Algorithms(VQAs)\cite{Kassal2011} and many VQAs are released such as Variational Quantum Eigensonlver(VQE)\cite{McClean_2016}, Adaptive VQE\cite{2019NatCo..10.3007G}, Multiscale Contracted VQE(MCVQE)\cite{2019arXiv190608728P}, and Variational Quantum Machine Learning Algorithms\cite{2019QS&T....4a4001K}\cite{2019Natur.567..209H}\cite{2021PhRvP..16d4057B}\cite{2022PhRvA.106b2601A}\cite{2020PhRvL.125j0401W}\cite{2022arXiv220211200K}\cite{2022arXiv220608316Y}\cite{PhysRevA.98.032309}.  Born machine is one of  VQMLs proposed in 2018\cite{2018arXiv180404168L}. The objective of this algorithm is to derive the quantum circuits and variables that make the aimed distribution of probability. There are no constraints on the circuits. 

Therefore, we propose the method that derives Hamiltonians that propagate initial states to aimed given states. The circuit is the propagator of the Hamiltonian. This method can be used to simulate the operations on quantum computers. Optimizing the architecture and compressing gate operations into one operation is also possible. We call this Method Hamiltonian Engineering by Born Machine (HEBM).
This is a generalized method of Quantum Coherent Ising Born Machine(QCIBM) \cite{coyle_born_2020} that can deal with any type of Hamiltonians.
QCIBM can only treat Ising model and its family.

 We derived some Hamiltonians for some pairs of initial and final distributions after propagation by HEBM. As a result, we confirmed that HEBM can derive the Hamiltonians that propagate initial states to aimed given states rapidly, and the accuracy is enough to be used practically. 

The organization of this paper is as follows. Chapter \ref{2} describes the details of HEBM. Chapter \ref{3} and \ref{4} indicate the result of our calculations. 

Chapter \ref{4} is a discussion for HEBM.

Chapter \ref{4.5} is the result and discussion of noise simulation.

Chapter \ref{5} is the conclusion of our work.

\section{Method}\label{2}

In this section, we describe the details of HEBM. The born machine is the Method in VQML. The objective of this algorithm is to derive the quantum circuits and variables that make the aimed distribution. The loss function of the Born machine uses kernels  to derive the quantum circuits and variables that make the aimed distribution like this, 

\begin{equation}
F = \bm{x} K \bm{x} + \bm{f} K \bm{f} - 2 \bm{f} K \bm{x}.
\end{equation}

This is the Maximum Mean Discrepancy Loss (MMD-Loss) function. Then, $\bm{x}$ and $\bm{f}$ are calculated distributions that are derived by the parametric quantum circuit and aimed distribution, respectively. They are $2^N$-dimensional vectors and expressed as $\bm{x} _j = \mid x_j \mid ^2 $ for $ \mid \Phi \rangle  = \sum_{j = 0}^{2^N} x_j \mid j \rangle$ and $\bm{ f }_j = \mid f_j \mid ^2 $ for $ \mid \Phi_{ ans } \rangle  = \sum_{j = 0}^{2^N} f_j \mid j \rangle$ by the decimal state represented by the state of each qubit. $K$ is the karnel matrix. $N$ is the number of qubits. The kernel is crucial for the Born machine. This is the matrix that interacts with the elements of distributions. Strictly speaking, it is the inner product of hyperfunctions that project variables into problem space. One can choose a given matrix element as that of a kernel matrix such as Gaussian, Poisson distribution, and SWAP-test. It is the archetype of VQAs. For example, it becomes Subspace Search VQE (SSVQE)\cite{2018arXiv181009434N} in case $\bm{x} = \bm{E_j}$ for the vector of trial energy $\bm{E_j} = \bm{\langle \Phi_j \mid H \mid \Phi_j \rangle}$. In case $j$ is only zero, and the cluster is made by the manner of Adaptive VQE, it is Adaptive VQE.
Circuits to arrange $\bm{x}$ chosen are clusters ordinarily. However,  the circuits for arranging $\bm{x}$ can be any type of circuit. We use the propagator of Hamiltonian, and the propagator is expressed as $exp(- i / 2 H t / N_{dt})^{N_{dt}}$ for $t = \pi / 2$. The parameters are coefficients of Hamiltonians, and Hamiltonians are expressed in Pauli's words. Then, $N_{dt}$ is the number of time frames. Hamiltonian takes the form $H = \sum_{j = 0}^{ N_o} \theta_j P_j$ for the product of the Pauli matrix $P_j$, consisted into Pauli matrix $X_j, Y_j, Z_j$.  $ N_o $ is the number of $P_j$ in Hamiltonian.

They are $2^N$-dimensional vectors and expressed as $\bm{x} = \sum_{j = 0}^{2^N} x_j \mid j \rangle$ and $\bm{ f } = \sum_{j = 0}^{2^N} f_j \mid j \rangle$ by the decimal state represented by the state of each qubit. 

Hence, the propagator must be decomposed into propagators of single terms of Hamiltonian. It is achieved by Suzuki-Trotter decomposition\cite{2017arXiv170102691R} in this work.

\begin{figure*}
\includegraphics[scale=0.3]{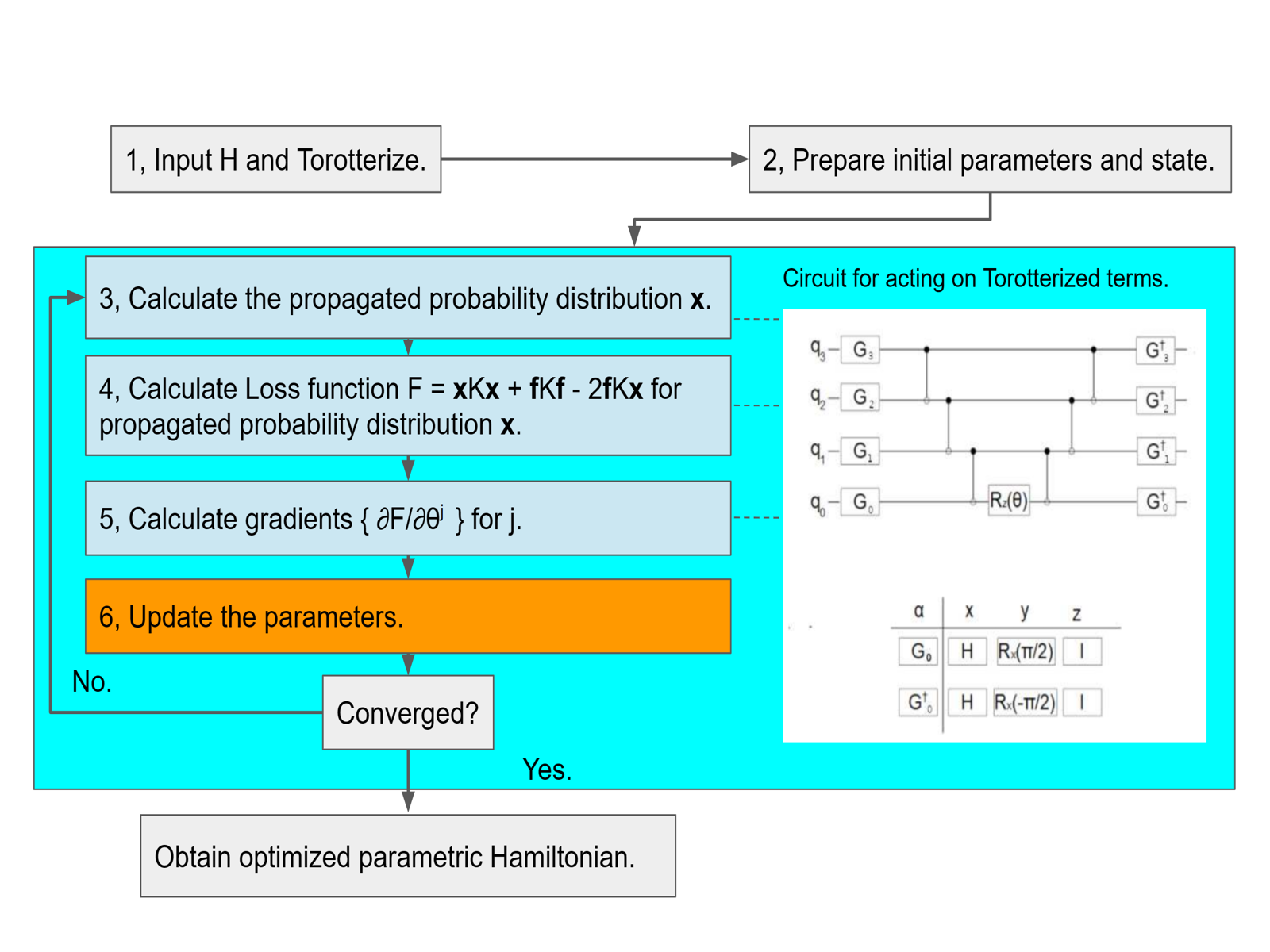}
\caption{
The flowchart of HEBM. 1, Firstly, parametric Hamiltonian H is input and Trotterized. 2, The initial parameters are set on H, and the initial circuit  $ \mid \Phi_{ ini } \rangle $  is prepared. 3, The initial State is propagated by H using the depicted circuit by $t = \pi / 2$ . 4, Loss function F is	calculated for the final calculated distribution $ \mid \Phi \rangle = exp(-iHt  / 2 ) \mid \Phi_{ ini } \rangle $. 5, The gradients of Loss function are calculated complying with parameter shift rule  ( PSR). 6, The parameters are updated by gradients. If the conditions of convergence are satisfied,  final parameters and Hamiltonian are obtained.
Otherwise, 3, 4, 5, and 6 are performed using updated parameters.}\label{HEBM}
\end{figure*}

In this method, the Born machine is used for Hamiltonian engineering. This is why we call this Method HEBM ( Figs.  \ref{HEBM} ).

We use L-Broyeden-Fletcher-Goldfarb-Shanno-B (L-BFGS-B) method\cite{doi:10.1002/9781118723203.ch3} for optimization of HEBM. The simulation of quantum simulation calculation is performed by blueqat SDK, the software development kit for quantum simulation and computation on Python. The number of shots is infinity; hence the result is read as a state vector.

\section{result}\label{3}

In this section, we perform to generate the optimal Hamiltonians that propagate initial states into aimed distributions. We perform this on Bars and Stripes distribution, Gaussian distribution, and Gibbs state for $H = - \sum_{j = 0}Z_j Z_{j + 1}$. The number of qubits is 4 for all cases. 

\begin{figure}
\includegraphics[scale=0.3]{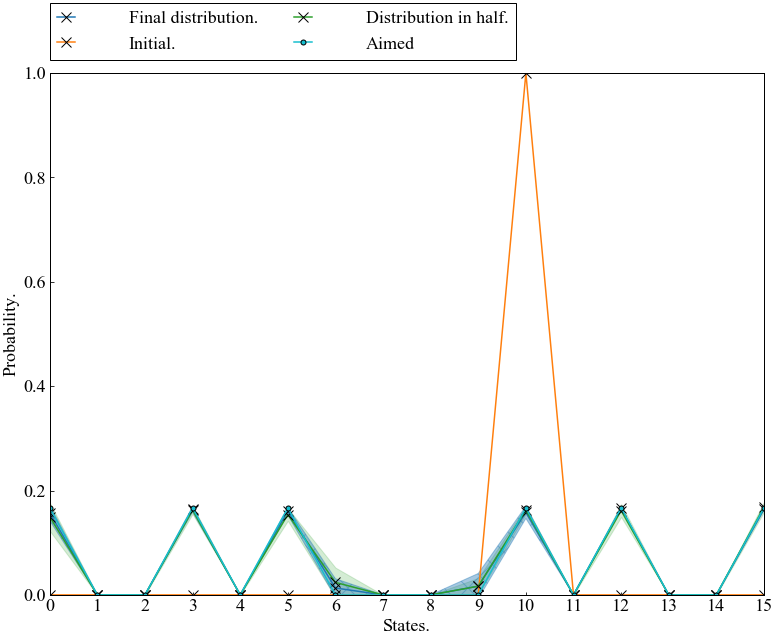}
\caption{
The number of states v.s. the average of probabilities of each state at the end and half of the total number of iterations for BAS distribution and their standard deviations. The standard deviations are the transplant areas of each line. }\label{basd}
\end{figure}

\begin{figure}
\includegraphics[scale=0.3]{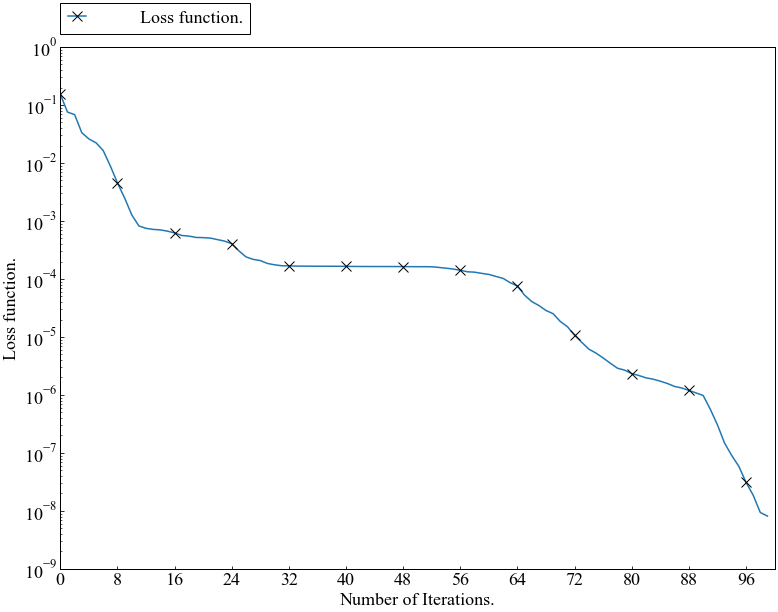}
\caption{
The number of iterations v.s. the loss function for BAS distribution. }\label{basf}
\end{figure}

Firstly, we perform the calculation on BAS distribution\cite{benedetti_generative_2019}. BAS distribution is the distribution that each qubit corresponds to the block on 2 $\times$ 2 tiles, and in case 0 and 1 make the row or column, they have a probability. Hence, 0, 3, 5, 10, 12, and 15 are $1 / 6$, and others are zero. We set the initial state $\mid 1010 \rangle = \mid 10 \rangle$. Hamiltonian is $H = \sum_{j = 0}^3 (\theta_{j}^{XX} X_j X_{j + 1} + \theta_{j}^{YY} Y_j Y_{j + 1} + \theta_{j}^{ZZ} Z_j Z_{1 + 1} + \theta_j^{Z} Z_j)$ for $j = 4 = 0$ (cyclic condition). We show the average of distributions of five iterations and the loss function of the fifth iteration in Fig. \ref{basd} and \ref{basf}, respectively. The average is almost the value of the aimed state. However, standard deviations of $\mid 7 \rangle$ and $\mid 8 \rangle$ are large. Rather those of distribution at the end are larger than those in half of the total number of iterations. Even though the loss function of the fifth iteration became a $10^{ - 8 }$ order and aimed distribution is realized, those of 3 of 5 iterations converged above $10^{- 4}$. the Hamiltonian that propagates $\mid 5 \rangle$ to BAS distribution is derived roughly. 


Second,  we perform the calculation on Gaussian distribution $e^{-  ( \bm{x} - c)^2 / 2 \sigma }$ . The center of distribution $c$ is $7.5$, and deviation $\sigma$ is 1. The initial state is equally distributed. 

Hamiltonian is  $H = \sum_{j = 0}^3 (\theta_{j}^{ZZ} Z_j Z_{1 + 1} + \theta_{j}^{ZXZ} Z_{j - 1} X_j Z_{1 + 1} + \theta_j^{X} X_j)$ that is the parent Hamiltonian of $H = - \sum_{j = 0}Z_j Z_{j + 1}$ for $j = -1 = 4 = 0$.

\begin{figure}
\includegraphics[scale=0.3]{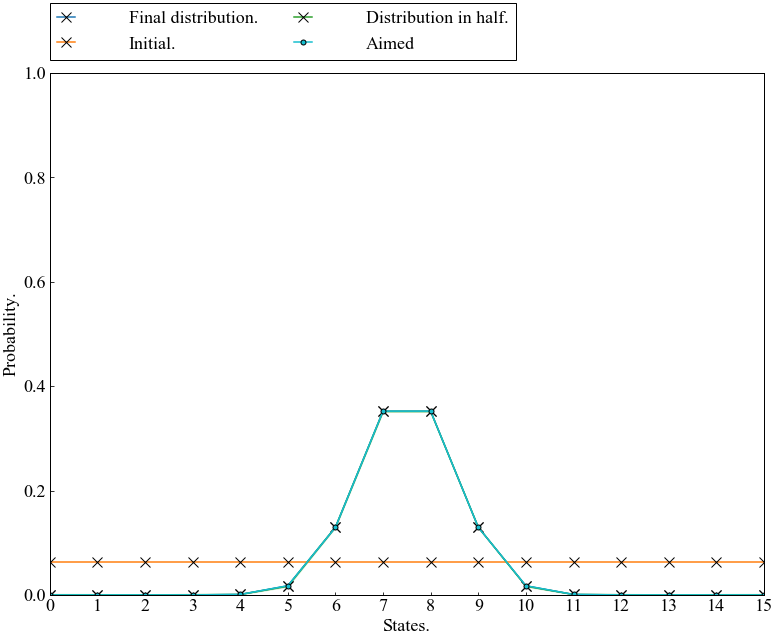}
\caption{
The number of states v.s. the average of probabilities of each state at the end and half of the total number of iterations for Gaussian distribution and their standard deviations. }\label{gad}
\end{figure}

\begin{figure}
\includegraphics[scale=0.3]{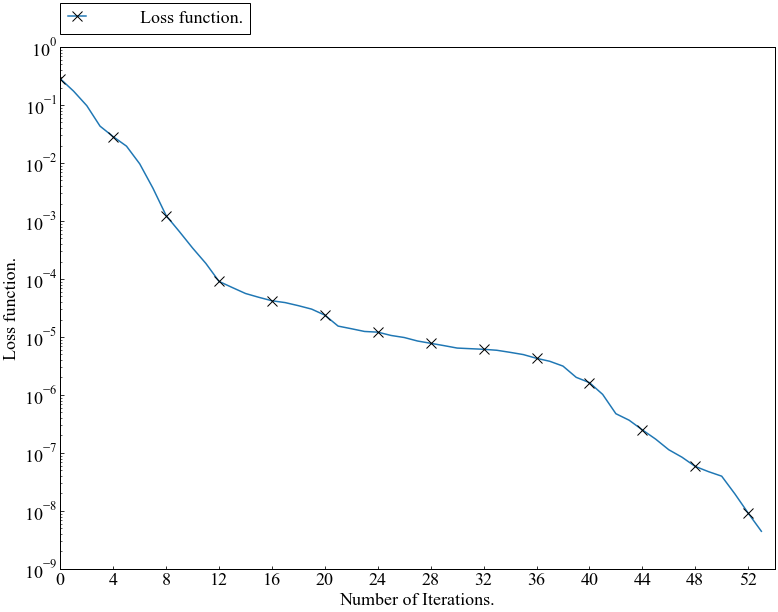}
\caption{
The number of iterations v.s. the loss function for Gaussian distribution. }\label{gaf}
\end{figure}

We show the average of distributions of five iterations and the loss function of the fifth iteration in Fig. \ref{gad} and \ref{gaf}, respectively. Both the averages of probabilities and their bounds of standard deviations have almost the same values as exact. loss function became $10^{- 9 }$ order and aimed distribution is realized. Hence, the Hamiltonian that propagates the equally distributed states to Gaussian is derived.


\begin{figure}
\includegraphics[scale=0.3]{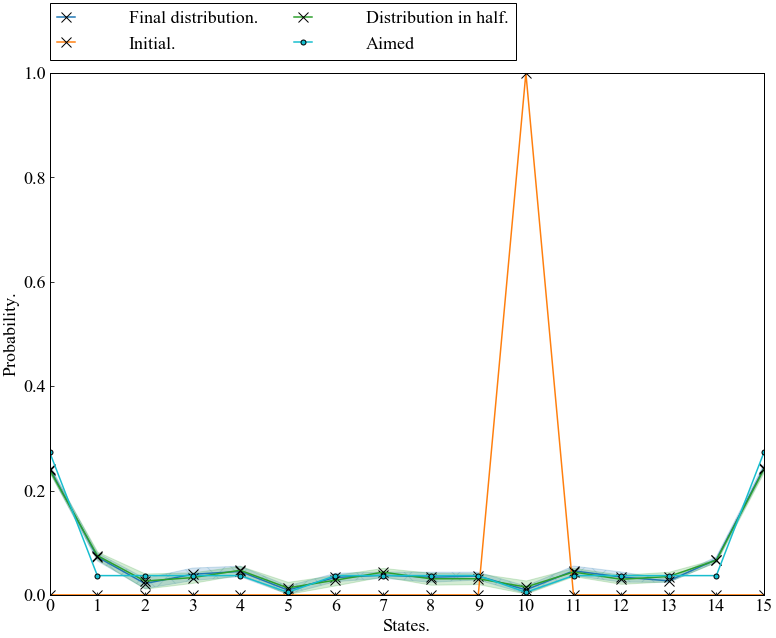}
\caption{
The number of states v.s. the average of probabilities of each state at the end and half of the total number of iterations for Gibbs distribution and their standard deviations in case the initial state is  $\mid 1010 \rangle = \mid 10 \rangle$ states. }\label{gibd}
\end{figure}

\begin{figure}
\includegraphics[scale=0.3]{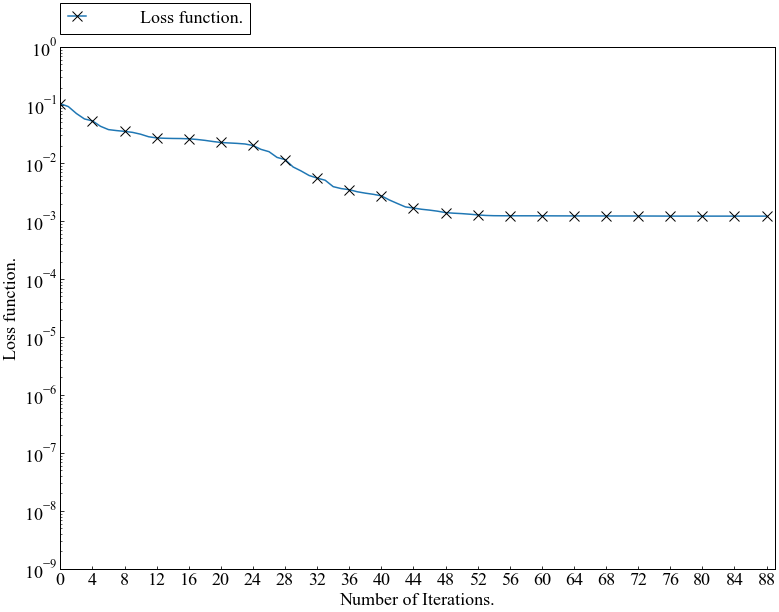}
\caption{
The number of iterations v.s. loss function for Gibbs distribution in case initial state is  $\mid 1010 \rangle = \mid 10 \rangle$ state. }\label{gibf}
\end{figure}

On the next hand, we calculated the Gibbs distribution\cite{2021arXiv211214688S} of  $H = - \sum_{j = 0}Z_j Z_{j + 1}$. Hamiltonian is the same as Gaussian. We show the average of distributions of five iterations and the loss function of the fifth iteration in case the initial state is  $\mid 1010 \rangle = \mid 10 \rangle$ state in Fig. \ref{gibd} and \ref{gibf}, respectively. The probability of $\mid 0 \rangle$, $\mid 1 \rangle$, $\mid 14 \rangle$, and $\mid 15 \rangle $ states deviate about 0.05 from those of aimed states. The standard deviations of other states are broad. loss function became $10^{- 4}$ order and aimed distribution is realized. Hence, the Hamiltonian propagates the given decimal states to which Gibbs distribution is difficult to be derived.

\begin{figure}
\includegraphics[scale=0.3]{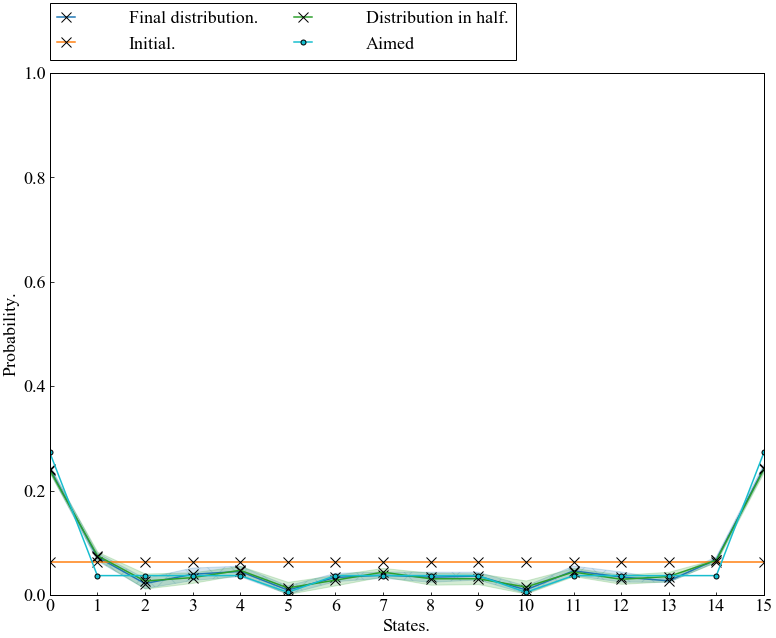}
\caption{
The number of states v.s. the average of probabilities of each state at the end and half of the total number of iterations for Gibbs distribution and their standard deviations in case the initial state is equally distributed. }\label{gibdh}
\end{figure}

\begin{figure}
\includegraphics[scale=0.3]{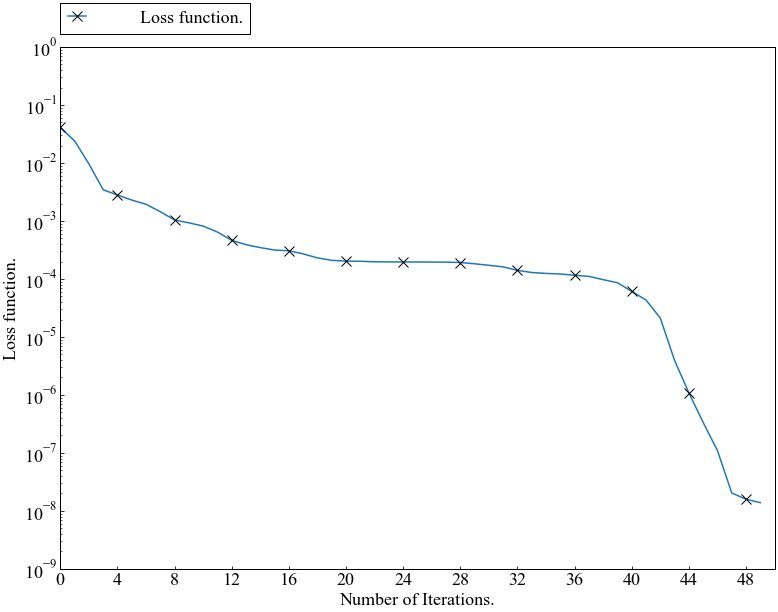}
\caption{
The number of iterations v.s. the loss function for Gibbs distribution in case the initial state is equally distributed. }\label{gibfh}
\end{figure}

On the other hand, the average of distributions of five iterations and loss function of the fifth iteration in case the initial state is equal distribution are in Figs. \ref{gibdh} and \ref{gibfh}, respectively. The standard deviations of $\mid 1 \rangle$, $\mid 2 \rangle$, $\mid 13 \rangle$, and $\mid 14 \rangle $ states deviate about 0.02  from those of aimed states. loss function became $10^{- 8 }$ order and aimed distribution is realized. Hence, the Hamiltonian propagates the equally distributed states to which Gibbs distribution is derived.



From these results, it is possible to derive the Hamiltonians that propagate the nonlocal distribution to nonlocal distributions and local distributions to local distributions. However, nonlocal distributions to local distributions and vice versa cannot be derived. In case we make a BAS distribution from an equal distribution,  the loss function cannot be below $10^{- 2}$. Besides, calculated distributions at the half of convergence sufficiently reproduce the aimed distributions. 

\section{discussion}\label{4}

\begin{figure}
\includegraphics[scale=0.3]{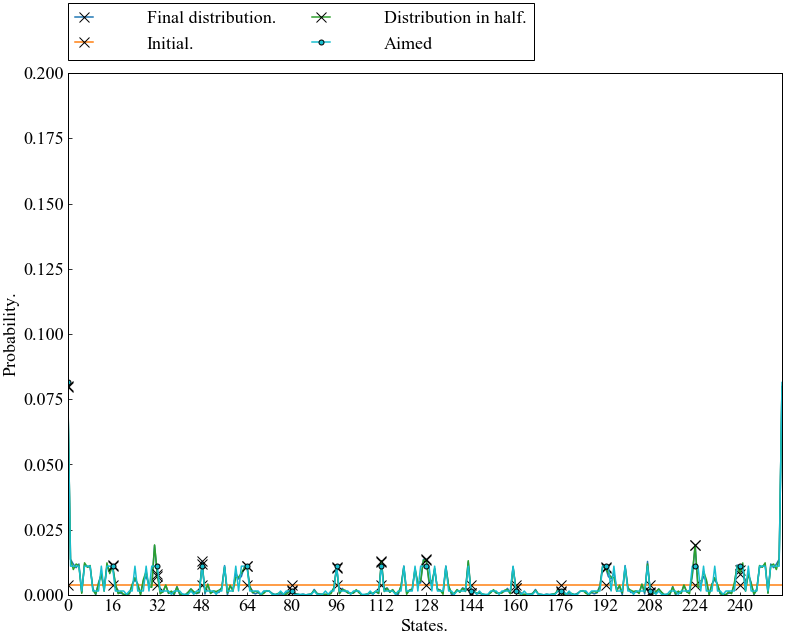}
\caption{
The number of state v.s. probabilities of each state at the end and half of the total number of iterations for the Gibbs distribution of the 8-qubit system. }\label{gi8}
\end{figure}

\begin{figure}
\includegraphics[scale=0.3]{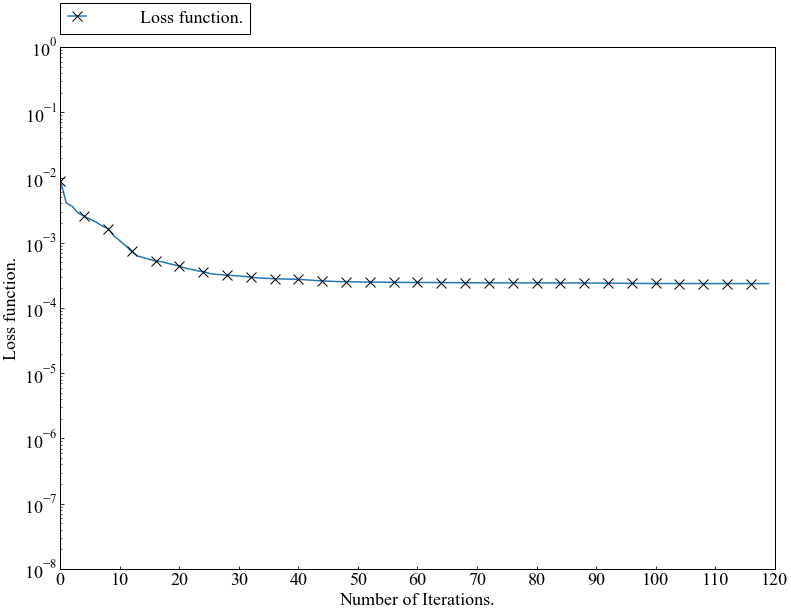}
\caption{
The number of iterations v.s. the loss function for Gibbs distribution of the 8-qubit system. }\label{gi8f}
\end{figure}

In this section, we discuss the advantages and shortcomings of our methods with data. The Hamiltonians that propagate initially to aimed given states can be derived rapidly. However, the more time frames  $N_{dt}$ be, the more time for one iteration will be taken. Therefore, we sampled the time for calculation for each $H_{dt}$ performing the calculation on Gaussian distribution. We show the result in Fig. ref{time}. The time for calculation is defined to be the time that the accuracy loss function becomes $<  10^{- 4}$. Time for calculation increases rapidly from $N_{dt} > 20$ and varies in the order of $O(atan N_{dt} + N_{dt})$. It indicates that our method is not sufficient for long simulations of quantum calculation algorithms. Though, some techniques may save time for simulations. For example, some processes in the simulation are compressed into one Hamiltonian. Some algorithms include repetition of unit processes such as Grover's algorithm. Hence, compressing the unit process may contribute to our method. The effects of the number of qubits are studied. We calculated the Gibbs distribution of  $H = - \sum_{j = 0}Z_j Z_{j + 1}$ for 8-qubit systems. According to Figs. \ref{gi8} and \ref{gi8f}, the final distribution is a little deviated from the aimed one, and the loss function is above $10^{- 5}$. Hence,  the accuracy declines a little for large systems. Besides, the time for calculation becomes longer as the number of qubits becomes larger. This can be avoided by techniques such as the density matrix renormalization group.

\begin{figure}
\includegraphics[scale=0.3]{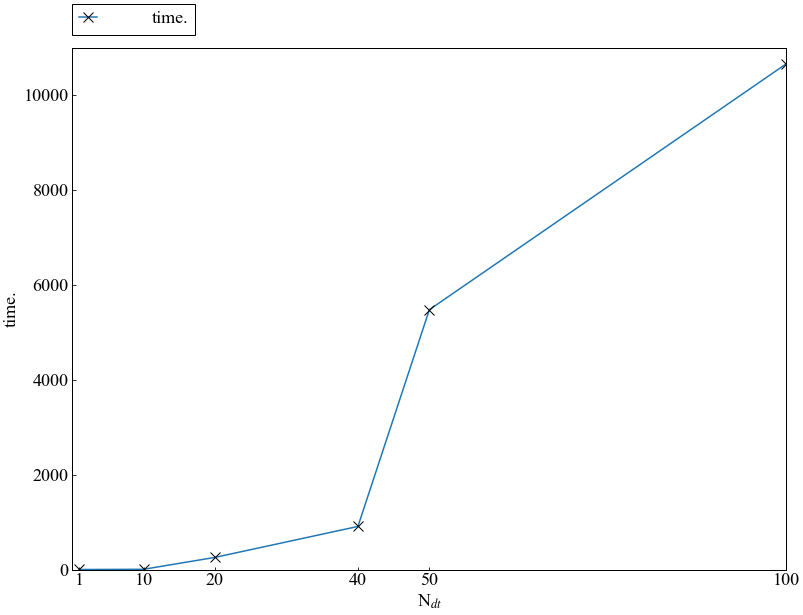}
\caption{
$N_{dt}$ v.s. time for calculations for Gaussian distribution. The unit is second for time. }\label{time}
\end{figure}


\section{Effect of noises}\label{4.5}

We simulate the noises by the reservoir by the scheme on paper \cite{2021PhRvL.127b0504H}.

\subsection{method of simulation of noises}

We describe the details of noisy simulation at first.

 Time propagation on a noisy system is depicted by the Lindblad equation shown below.

\begin{eqnarray}
\frac{ d \rho } { d t }   &=&   - \sum_j i [ P_j ,   \rho ] +  \sum_k  (  2  L_k^{  \dagger }  \rho L_k  - [ L_k^{  \dagger } L_k  , \rho  ]_+  ) \\ 
 &=& \sum_j  \mathscr{ E }_j +  \sum_k  \mathscr{ L }_k  \\ 
 \mathscr{ E }_j   &=&   - i [ P_j ,   \rho ]   \\ 
\mathscr{ L }_k &=&   2  L_k^{  \dagger }  \rho L_k  - [ L_k ^{  \dagger } L_k  , \rho  ]_+    \\ \nonumber
\end{eqnarray}

Then, $\rho$ is the density matrix of the system, and $ P_j  $ is the j -th term of the Hamiltonian, $L_k$ is the Lindbradian of k-th noise term, respectively. 
There are two main noises: dephasing phase kick and pole kick, respectively. 
The prior is the relaxation of phase described by $L = \frac{ \sqrt{ \gamma_1 }  }{  2  }  Z$   in Lindbradian. 
The posterior is the relaxation of amplitude described by $L = { \sqrt{ \gamma_2  }  } ( X - i Y )$   in Lindbradian. 
The propagator of the Lindblad equation is, 

\begin{equation}
\mathscr{  U  } = e^{ ( \sum_j  \mathscr{ E }_j +  \sum _k  \mathscr{ L }_k )t 
 } .
\end{equation}

The propagator of the Lindblad equation can also be Trotterized as,

\begin{equation}
 lim_{ N \rightarrow  \infty } ( \prod_ j e^ { \mathscr{ E }_j  t / N  }  \prod_k   e^{  \mathscr{ L }_k t/ N   } )  = e^{ ( \sum_ j  \mathscr{ E }_j +  \sum _k  \mathscr{ L }_k )t }  .
\end{equation}

\begin{figure*}

\includegraphics[scale=0.3]{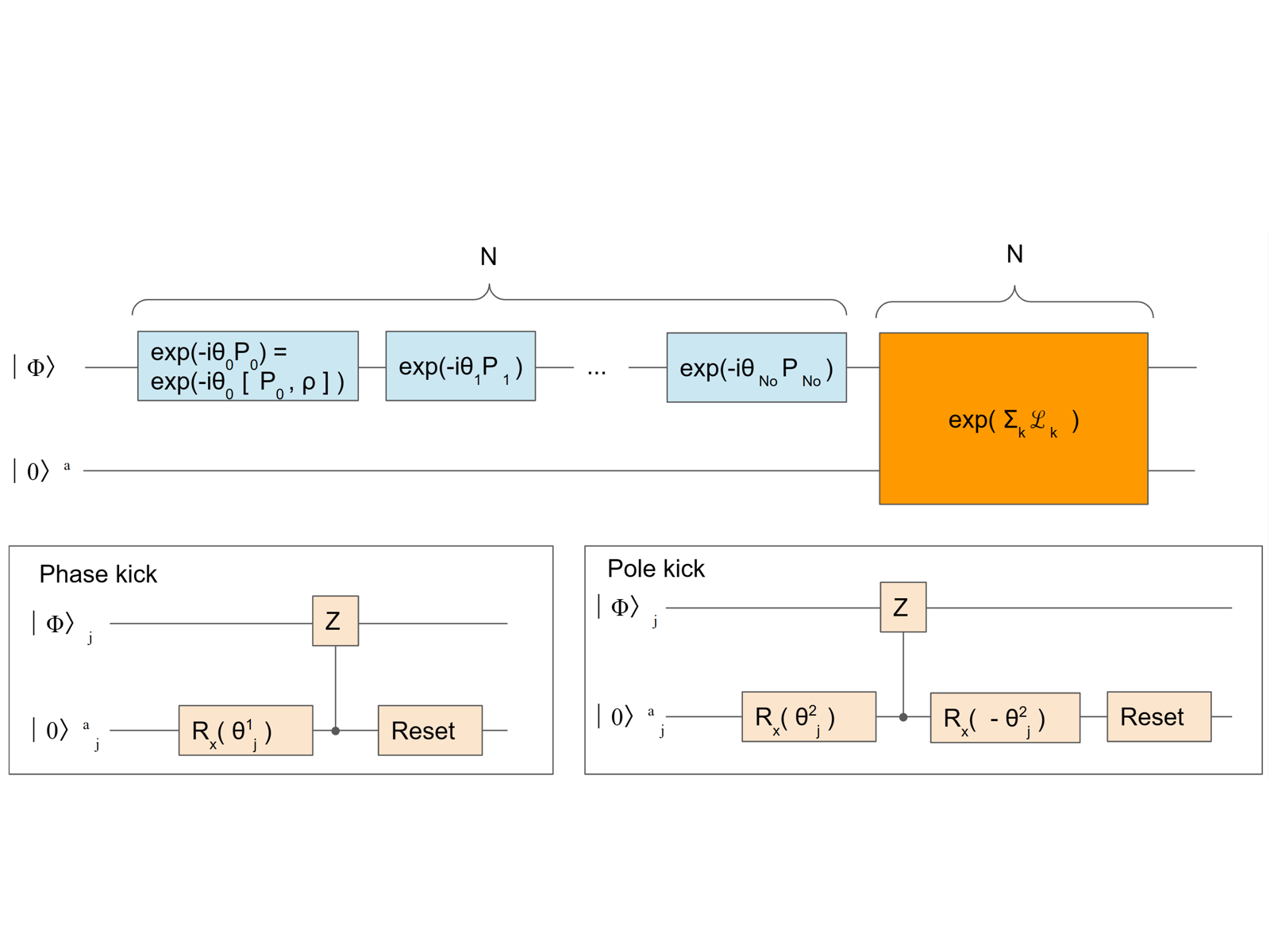}
\caption{
The quantum circuit to implement the propagator of HEBM in a noisy environment. 
The time propagator consists of the Hamiltonian part( blue ) and the noise part  ( orange ). 
The noise part contains phase kick and pole kick, illustrated below. 
The Reset gate makes the state of qubit 0 states.
}\label{noises}

\end{figure*}

$ \gamma_1 $ and $ \gamma_2 $ are jumping rates expressed as  $ \gamma_1 = - ln ( 2 cos^2   \theta^1 - 1 ) / \tau_0 $  and $ \gamma_2 = - ln (  cos^2   \theta^2   ) / \tau_0 $ , respectively. 
$ \tau_0 $  is the period of noise, described by  $ \tau_0 = e_{ unit. } /  \hbar N_{ d t } $  for this paper where $ e_{ unit. } $ is the unit of energy in the SI  unit. 
This propagator can also be implemented the same as an ordinary noiseless time propagator. 
Fig. \ref{noises} depicts the quantum circuit to implement the propagator of the Lindblad equation on unit frame time.
 In this paper, the effect of noise is simplified into the noises for each qubit; hence, each qubit is acted noise simulation by corresponding ancilla qubit. 
Law noisy simulation requires 240 variable parameters for only two qubits.

  We assume the angle is the random value that the maximum absolute value is manually given noise variance. $ N $ is 1 for Hamiltonian because there is no difference between the accuracy of whether 1 or 13. $ N $ is 13 for all simulations of noises and we use second-order Trotterization.

\subsection{ simulation of noises}

\begin{figure}

\includegraphics[scale=0.3]{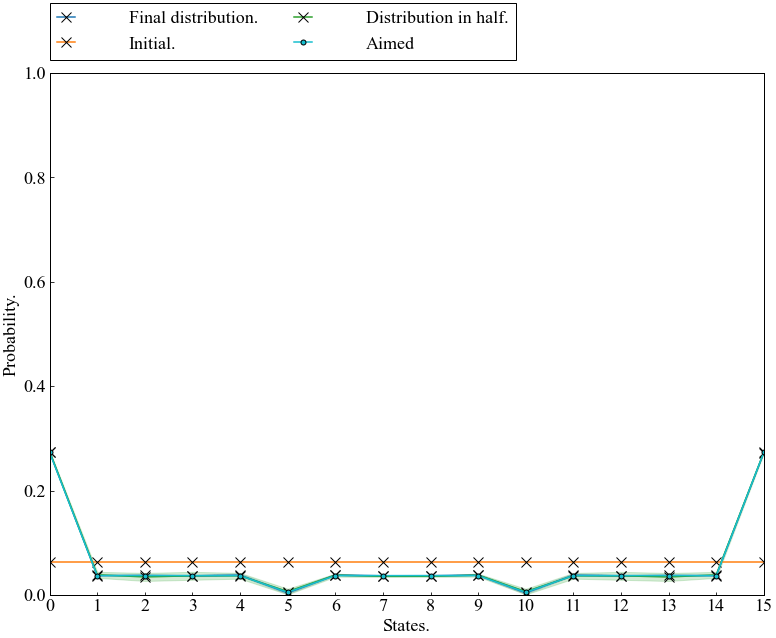}

\caption{The number of states v.s. the average of probabilities of each state at the end and half of the total number of iterations for Gibbs distribution and their standard deviations when the noise variances in angular of phase and pole kicks are both 5. }\label{g5d}

\end{figure}

\begin{figure}

\includegraphics[scale=0.3]{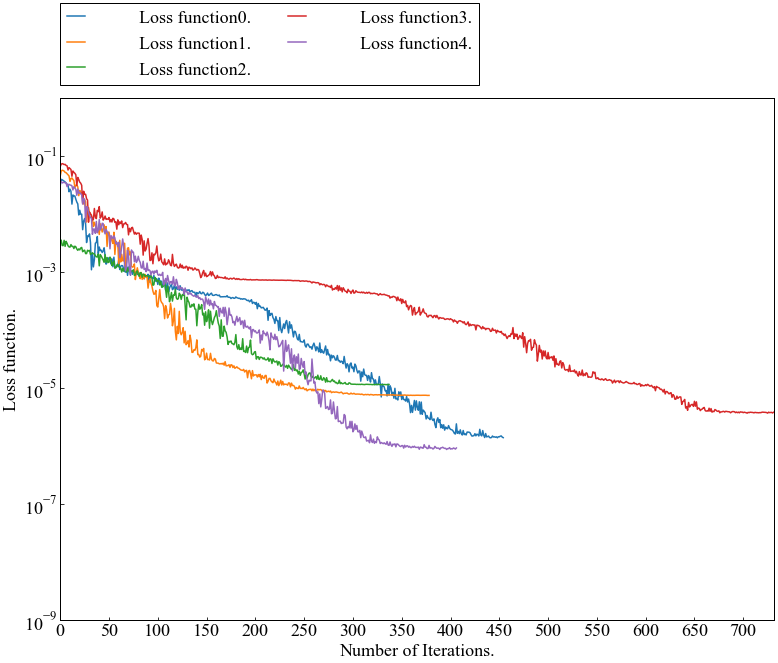}

\caption{The number of iterations v.s. the loss function for Gibbs distribution in case the initial state is an equally distributed state when the noise variances in angular of phase and pole kicks are both 5. }\label{5f}

\end{figure}

\begin{figure}

\includegraphics[scale=0.3]{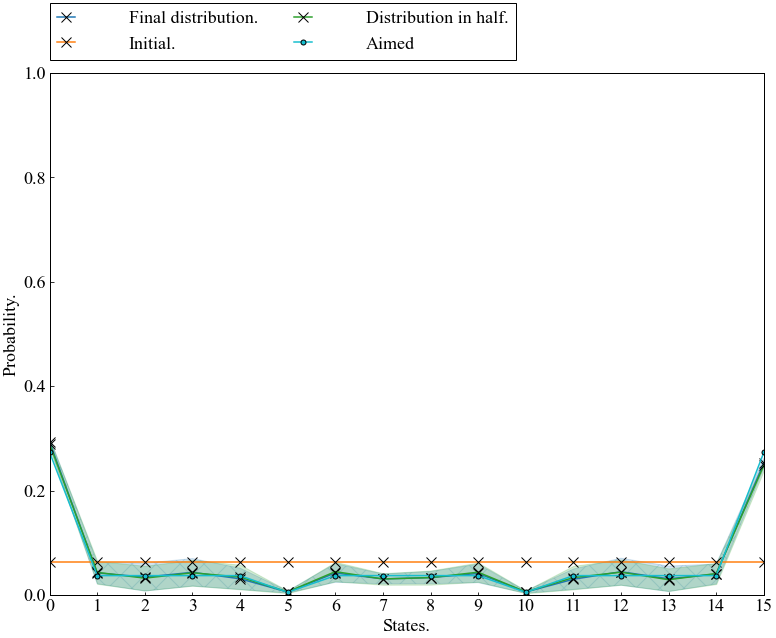}

\caption{The number of states v.s. the average of probabilities of each state at the end and half of the total number of iterations for Gibbs distribution and their standard deviations when the noise variances in angular of phase and pole kicks are both 30. }\label{g30d}

\end{figure}

\begin{figure}

\includegraphics[scale=0.3]{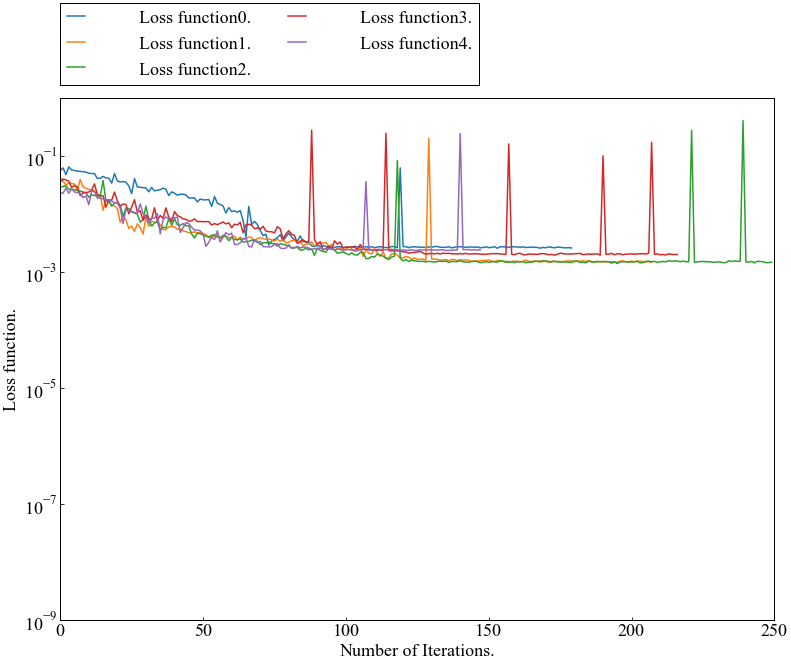}

\caption{The number of iterations v.s. the loss function for Gibbs distribution in case the initial state is an equally distributed state when the noise variances in angular of phase and pole kicks are both 30. }\label{30f}

\end{figure}

First, we show the result of the quantum Gibbs distribution. 
The initial state is equally distributed. The calculated distributions at the half and the end of iterations when the noise divergences in angular of phase and pole kick are both 5 and 30 in Figs. \ref{g5d} and \ref{g30d}, respectively.  The standard deviations of distributions when noise divergences are both 30 are larger than that when the noise variances are both 5, even though the average distributions are nearly exact for both cases. The convergence of loss functions showed more differences. We show the convergence of five samples of loss functions when noise variances are both 5 and both 30  degrees in Figs.  \ref{5f} and  \ref{30f}, respectively. All samples converged into $ 10 ^{- 2 } $ order when noise variances are both 30. In contrast, all samples converged less than $ 10 ^{-4 } $ order when noise variances are both 5 degrees.

\begin{figure}

\includegraphics[scale=0.3]{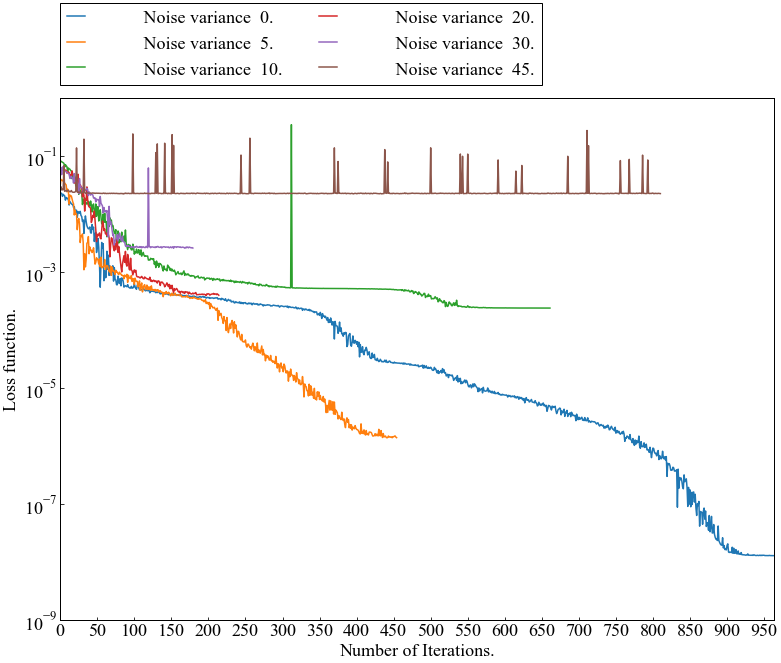}

\caption{The number of iterations v.s. 0-th sample of logarithmic loss function for Gibbs distribution in case the initial state is an equally distributed state when the noise variances in angular of phase and pole kicks are both 0, 5, 10, 20, 30, and 45 degrees, respectively.}\label{afang}

\end{figure}

The accuracy of calculation by HEBM declines drastically as the noise variance increases. Fig. \ref{afang} shows the convergences of zeroth samples when noise variances are both 0, 5, 10, 20, 30, and 45 degrees, respectively. The loss functions when noise variances are both more than 10 are all converged into more than $ 10 ^{-4 } $ even other ones when noise variances are both 0   and 5 converged less than $ 10 ^{ - 5 } $. Those when noise variances are both 30 and 45 have not converged less than $ 10^{ -3} $. The plateaus of both are supposed to be because noises nullify the effect of optimizers.

 In addition, The result of HEBM has higher accuracy than the case that Hamiltonian is Ising Hamiltonian(QCIBM) for all noise variance.


\begin{figure}

\includegraphics[scale=0.15]{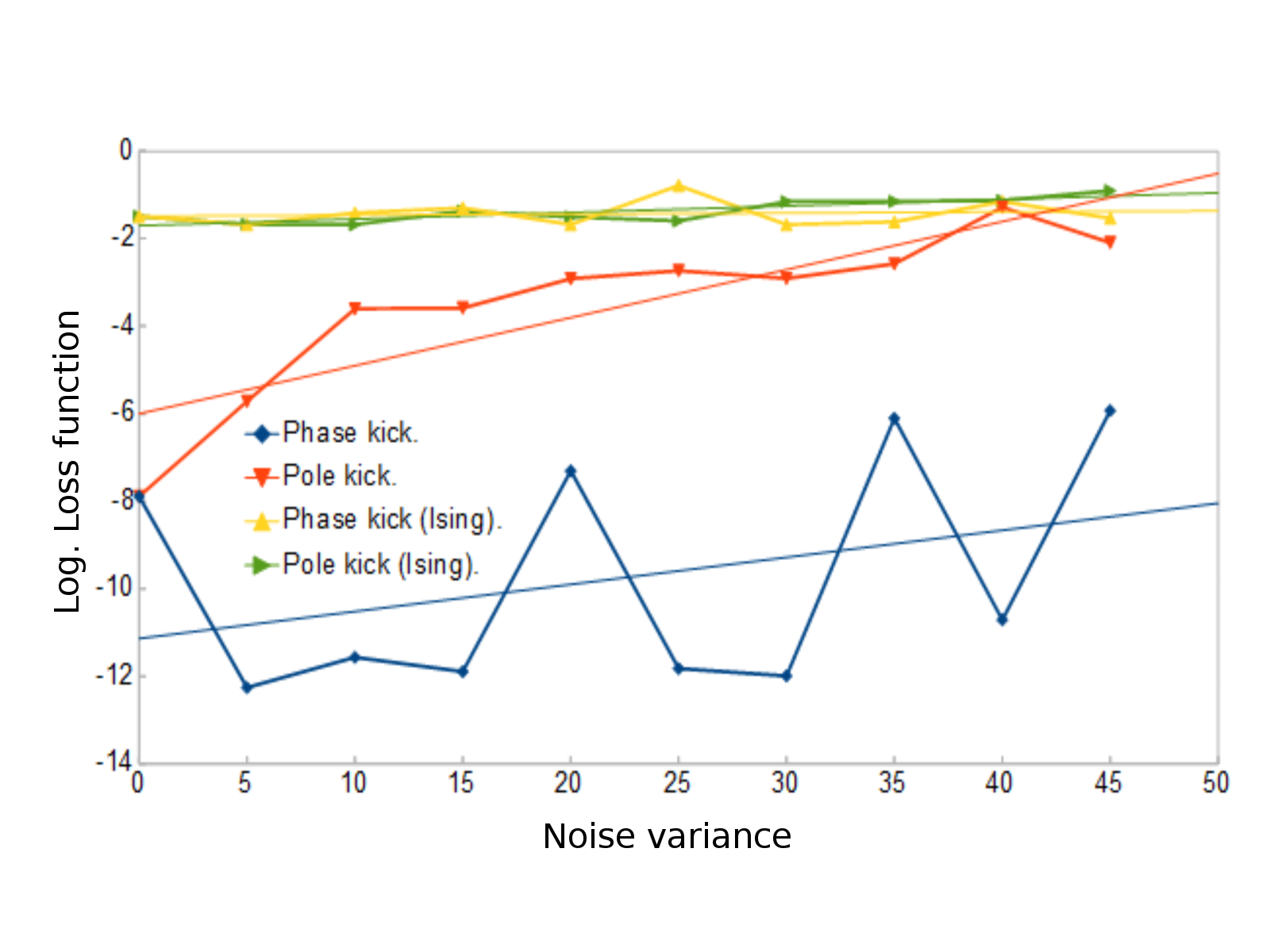}

\caption{The noise variances in angular of phase and pole kicks v.s. the loss function for Gibbs distribution in case the initial state is equally distributed. }\label{gab}

\end{figure}

We also investigated the effect of both phase kick and pole kick noises. We calculated the final loss for every 5 degrees from 0 to 45 degrees and sampled once on both phase and pole kicks. We show the result of sampling in Fig.  \ref{gab}. There is no doubt that polar hick is mainly affecting the accuracy. On the other hand, phase kick lowers the accuracies only two digits when the noise variance is 45 degrees. However, it is unstable. 

\begin{figure}

\includegraphics[scale=0.3]{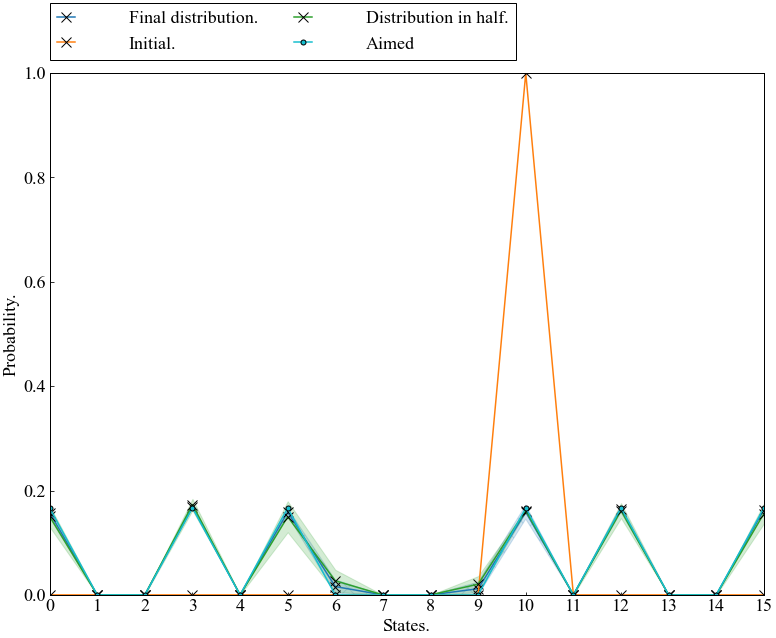}

\caption{The number of states v.s. the average of probabilities of each state at the end and half of the total number of iterations for Gibbs distribution and their standard deviations when the noise variances in angular of phase and pole kicks are both 5. }\label{bas5d}

\end{figure}

\begin{figure}

\includegraphics[scale=0.3]{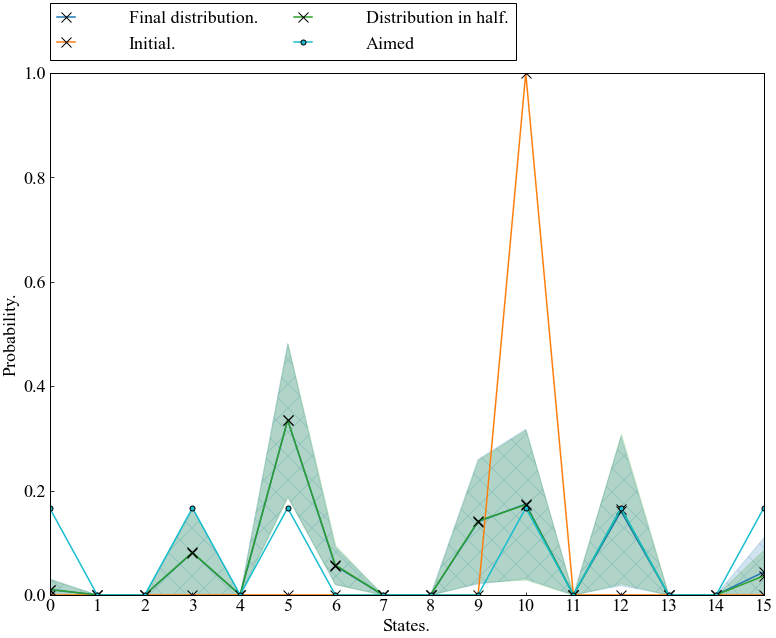}

\caption{The number of states v.s. the average of probabilities of each state at the end and half of the total number of iterations for BAS distribution and their standard deviations when the noise variances in angular of phase and pole kicks are both 30. }\label{bas30d}

\end{figure}

Secondly, we show the result of BAS distribution. The calculated distributions at the half and the end of iterations, when the noise variances in angular of phase and pole kicks are both 5 and 30 degrees, are shown in Figs. \ref{bas5d}  and \ref{bas30d}, respectively. 

Both the distributions at the end and half of the iteration are off exact values at  $ \mid 0 \rangle $,  $ \mid 3 \rangle $,  $ \mid 5 \rangle $,  $ \mid 9 \rangle $, and  $ \mid 15 \rangle $  states when house variances are both 30 degrees. 

\begin{figure}

\includegraphics[scale=0.3]{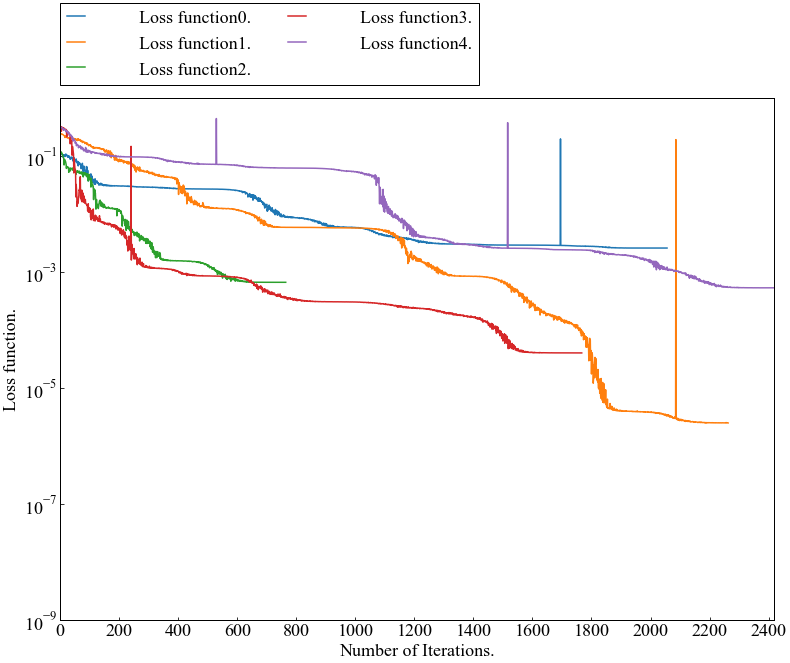}

\caption{The number of iterations v.s. the loss function for BAS distribution in case the initial state is an equally distributed state when the noise variances in angular of phase and pole kicks are both 5. }\label{bas5f}

\end{figure}

\begin{figure}

\includegraphics[scale=0.3]{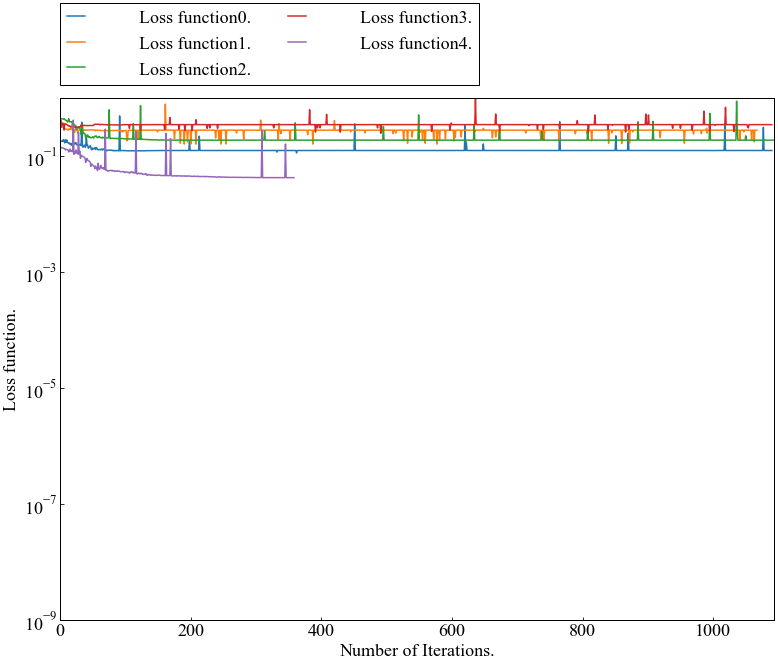}

\caption{The number of iterations v.s. the loss function for BAS distribution in case the initial state is an equally distributed state when the noise variances in angular of phase and pole kicks are both 30. }\label{bas30f}

\end{figure}

\begin{figure}

\includegraphics[scale=0.3]{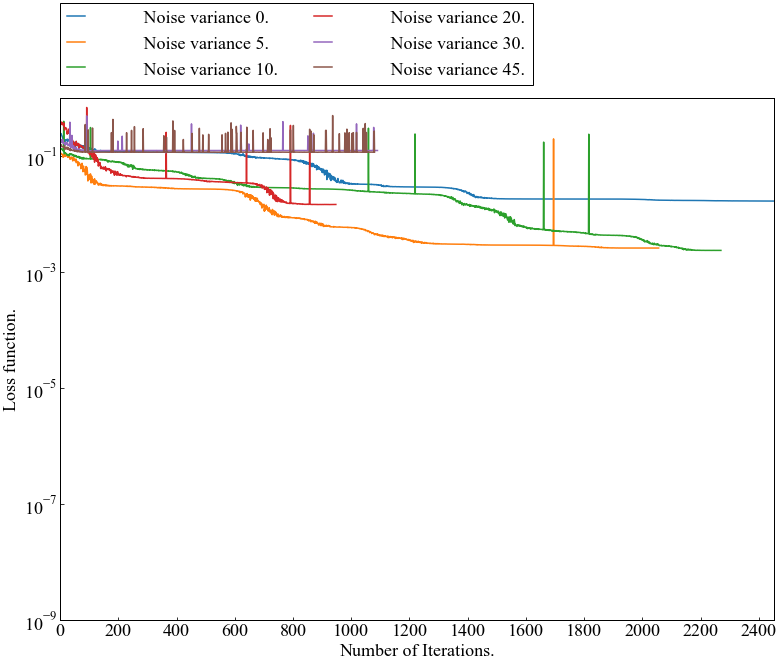}

\caption{The number of iterations v.s. 0-th sample of logarithmic loss function for BAS distribution in case the initial state is an equally distributed state when the noise variances in angular of phase and pole kicks are both 0, 5, 10, 20, 30, and 45 degrees, respectively. }\label{basfang}

\end{figure}

\begin{figure}

\includegraphics[scale=0.15]{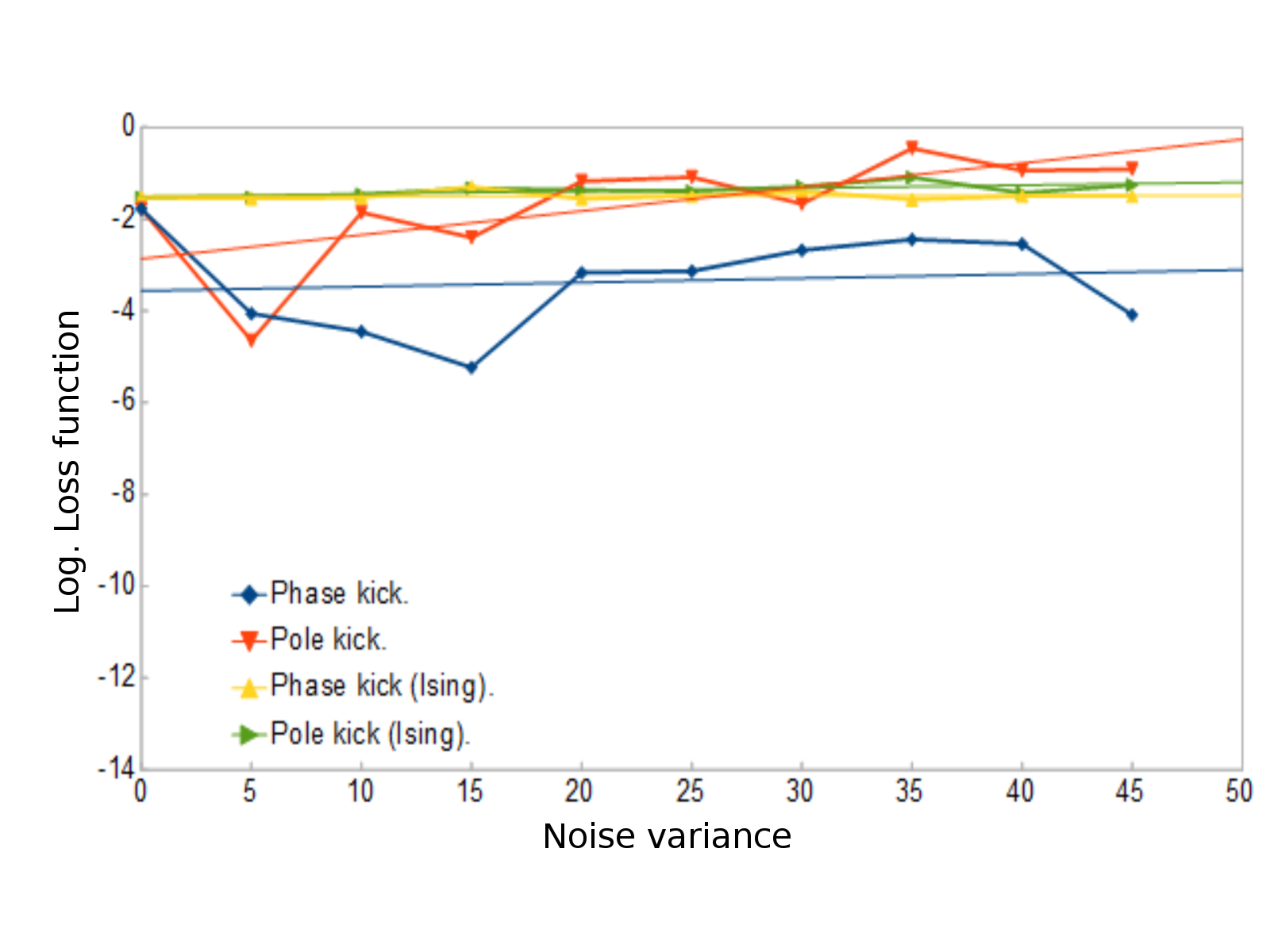}

\caption{The noise variances in angular of phase and pole kicks v.s. the loss function for BAS distribution in case the initial state is equally distributed. }\label{basab}

\end{figure}

The standard deviations are also off exact values and broad. In contrast, the calculated distributions when the noise variances are both 5 degrees are nearly exact values in almost all states. The convergence of loss functions is more variant than those of quantum Gibbs distribution.
We show the convergences of five samples of loss functions when the noise variances are both 5 and both 30 degrees in Figs.  \ref{bas5f}   and \ref{bas30f}, respectively. Only one sample converged less than $ 10^{ -5 } $  when the noise variances were 5 degrees. Besides, all samples converged into different orders. On the other hand, all samples converged more than $ 10^{  -2 } $ when the house variance was 30 degrees. Four of five samples have not converged due to noise, the same as the quantum Gibbs distribution. The noise variance declines, and The accuracy of calculation by HEBM declines drastically as the noise valiance too. Fig.  \ref{basfang} shows the convergence of zeroth samples when the noise variances are both  0, 5, 10, 20, 30, and 45, respectively. Samples converged unstably below $ 10^{    -1} $ when the noise variable was less than 30 degrees and did not converge when it was equal to or more than 30 degrees. We investigated the effect of phase and pole kick also for BAS distribution. Pole kick mainly affects the accuracy. We show the result of sampling in Fig.  \ref{basab}. Phase kick is negligible because bare accuracy on BAS distribution of HEBM is itself low.

  In addition, The result of HEBM has higher accuracy than the case that the Hamiltonian is Ising Hamiltonian (QCIBM) for noise variances below 30.


In this section, we describe the relationship among the accuracy of HEBM, KL divergence, and noise variances on the Gauss distribution. 

The initial state is an equally distributed state of 4 - qubit system, and sampled points are b = 0.1, 0.125, 0.15, 0.175, 0.2, 0.225, 0.25, 0.275, 0.3, 0.325, 0.35, 0.4, 0.5, 1.0, 2.0, 4.0, respectively. 

\begin{figure}

\includegraphics[scale=0.15]{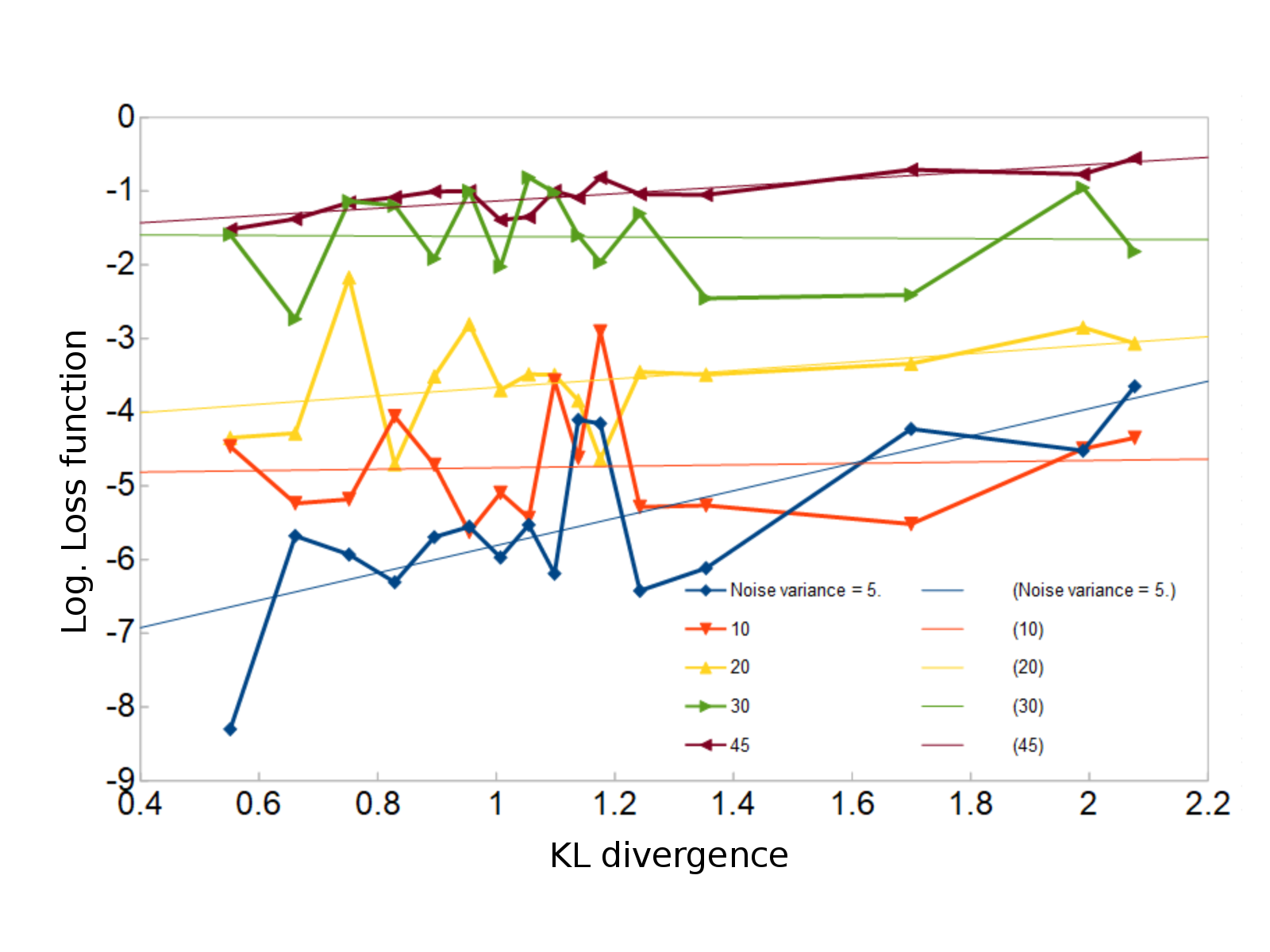}

\caption{The KL divergence v.s. the logarithmic loss function for Gauss distribution in case the initial state is an equally distributed state when the noise variances in angular of phase and pole kicks are both 0, 5, 10, 20, 30, and 45 degrees, respectively. }\label{gaukl}

\end{figure}

Each value of b corresponds to a given KL divergence. Fig.  \ref{gaukl}   shows the minimum logarithms Loss function in 5 samples for each KL divergence and noise variance. The minimum Loss functions are affected by KL divergence in case noise variance is 5 and 45. When noise variance is 5, the effect of noises is nearly negligible. On the other hand, the effect of noises is maximum in case the noise variance is 45. As KL divergence increases, the logarithm of the minimum Loss function rises gradually. According to the result on BAS distribution, KL divergence mainly affects the probability that the Loss function converges into minimum values for a noise variance. Besides, the noise variance mainly affects the accuracy. 

In case the noise variance is below 10, the logarithms of the minimum loss function are below - 4.

\section{Concluding remarks}\label{5}.

It was confirmed that the Hamiltonians that propagate initially to aimed given states can be derived by the Born machine.
This method can be used for hardware simulation of quantum computers. The robustness for dephasing and depolarizing is revealed, too. If depolarization is below 10 at a minimum, HEBM can derive the Hamiltonians accurately enough on read quantum devices. Benchmarking in real quantum devices is also a problem. However, some issues remained. As both the sampling rate and the number of qubits increases, the time for calculation increases. They can be solved by some techniques mentioned in the previous section. This method may contribute to not only hardware simulation but also fabricating materials in case this method can become able to be used for large systems.

\newpage\bibliographystyle{apsrev4-2}
\bibliography{ref_hebm}

\begin{thebibliography}{26}%
\makeatletter
\providecommand \@ifxundefined [1]{%
 \@ifx{#1\undefined}
}%
\providecommand \@ifnum [1]{%
 \ifnum #1\expandafter \@firstoftwo
 \else \expandafter \@secondoftwo
 \fi
}%
\providecommand \@ifx [1]{%
 \ifx #1\expandafter \@firstoftwo
 \else \expandafter \@secondoftwo
 \fi
}%
\providecommand \natexlab [1]{#1}%
\providecommand \enquote  [1]{``#1''}%
\providecommand \bibnamefont  [1]{#1}%
\providecommand \bibfnamefont [1]{#1}%
\providecommand \citenamefont [1]{#1}%
\providecommand \href@noop [0]{\@secondoftwo}%
\providecommand \href [0]{\begingroup \@sanitize@url \@href}%
\providecommand \@href[1]{\@@startlink{#1}\@@href}%
\providecommand \@@href[1]{\endgroup#1\@@endlink}%
\providecommand \@sanitize@url [0]{\catcode `\\12\catcode `\$12\catcode `\&12\catcode `\#12\catcode `\^12\catcode `\_12\catcode `\%12\relax}%
\providecommand \@@startlink[1]{}%
\providecommand \@@endlink[0]{}%
\providecommand \url  [0]{\begingroup\@sanitize@url \@url }%
\providecommand \@url [1]{\endgroup\@href {#1}{\urlprefix }}%
\providecommand \urlprefix  [0]{URL }%
\providecommand \Eprint [0]{\href }%
\providecommand \doibase [0]{https://doi.org/}%
\providecommand \selectlanguage [0]{\@gobble}%
\providecommand \bibinfo  [0]{\@secondoftwo}%
\providecommand \bibfield  [0]{\@secondoftwo}%
\providecommand \translation [1]{[#1]}%
\providecommand \BibitemOpen [0]{}%
\providecommand \bibitemStop [0]{}%
\providecommand \bibitemNoStop [0]{.\EOS\space}%
\providecommand \EOS [0]{\spacefactor3000\relax}%
\providecommand \BibitemShut  [1]{\csname bibitem#1\endcsname}%
\let\auto@bib@innerbib\@empty
\bibitem [{\citenamefont {Feynman}(1982)}]{feynman_simulating_1982}%
  \BibitemOpen
  \bibfield  {author} {\bibinfo {author} {\bibfnamefont {R.~P.}\ \bibnamefont {Feynman}},\ }\href {https://doi.org/10.1007/BF02650179} {\bibfield  {journal} {\bibinfo  {journal} {International Journal of Theoretical Physics}\ }\textbf {\bibinfo {volume} {21}},\ \bibinfo {pages} {467} (\bibinfo {year} {1982})}\BibitemShut {NoStop}%
\bibitem [{\citenamefont {{Lavor}}\ \emph {et~al.}(2003)\citenamefont {{Lavor}}, \citenamefont {{Manssur}},\ and\ \citenamefont {{Portugal}}}]{2003quant.ph..1079L}%
  \BibitemOpen
  \bibfield  {author} {\bibinfo {author} {\bibfnamefont {C.}~\bibnamefont {{Lavor}}}, \bibinfo {author} {\bibfnamefont {L.~R.~U.}\ \bibnamefont {{Manssur}}},\ and\ \bibinfo {author} {\bibfnamefont {R.}~\bibnamefont {{Portugal}}},\ }\href@noop {} {\bibfield  {journal} {\bibinfo  {journal} {arXiv e-prints}\ ,\ \bibinfo {eid} {quant-ph/0301079}} (\bibinfo {year} {2003})},\ \Eprint {https://arxiv.org/abs/quant-ph/0301079} {arXiv:quant-ph/0301079 [quant-ph]} \BibitemShut {NoStop}%
\bibitem [{\citenamefont {Shor}(1994)}]{365700}%
  \BibitemOpen
  \bibfield  {author} {\bibinfo {author} {\bibfnamefont {P.}~\bibnamefont {Shor}},\ }in\ \href {https://doi.org/10.1109/SFCS.1994.365700} {\emph {\bibinfo {booktitle} {Proceedings 35th Annual Symposium on Foundations of Computer Science}}}\ (\bibinfo {year} {1994})\ pp.\ \bibinfo {pages} {124--134}\BibitemShut {NoStop}%
\bibitem [{\citenamefont {Mezzacapo}\ \emph {et~al.}(2013)\citenamefont {Mezzacapo}, \citenamefont {Casanova}, \citenamefont {Lamata},\ and\ \citenamefont {Solano}}]{Mezzacapo_2013}%
  \BibitemOpen
  \bibfield  {author} {\bibinfo {author} {\bibfnamefont {A.}~\bibnamefont {Mezzacapo}}, \bibinfo {author} {\bibfnamefont {J.}~\bibnamefont {Casanova}}, \bibinfo {author} {\bibfnamefont {L.}~\bibnamefont {Lamata}},\ and\ \bibinfo {author} {\bibfnamefont {E.}~\bibnamefont {Solano}},\ }\href {https://doi.org/10.1088/1367-2630/15/3/033005} {\bibfield  {journal} {\bibinfo  {journal} {New Journal of Physics}\ }\textbf {\bibinfo {volume} {15}},\ \bibinfo {pages} {033005} (\bibinfo {year} {2013})}\BibitemShut {NoStop}%
\bibitem [{\citenamefont {Dutt}\ \emph {et~al.}(2007)\citenamefont {Dutt}, \citenamefont {Childress}, \citenamefont {Jiang}, \citenamefont {Togan}, \citenamefont {Maze}, \citenamefont {Jelezko}, \citenamefont {Zibrov}, \citenamefont {Hemmer},\ and\ \citenamefont {Lukin}}]{doi:10.1126/science.1139831}%
  \BibitemOpen
  \bibfield  {author} {\bibinfo {author} {\bibfnamefont {M.~V.~G.}\ \bibnamefont {Dutt}}, \bibinfo {author} {\bibfnamefont {L.}~\bibnamefont {Childress}}, \bibinfo {author} {\bibfnamefont {L.}~\bibnamefont {Jiang}}, \bibinfo {author} {\bibfnamefont {E.}~\bibnamefont {Togan}}, \bibinfo {author} {\bibfnamefont {J.}~\bibnamefont {Maze}}, \bibinfo {author} {\bibfnamefont {F.}~\bibnamefont {Jelezko}}, \bibinfo {author} {\bibfnamefont {A.~S.}\ \bibnamefont {Zibrov}}, \bibinfo {author} {\bibfnamefont {P.~R.}\ \bibnamefont {Hemmer}},\ and\ \bibinfo {author} {\bibfnamefont {M.~D.}\ \bibnamefont {Lukin}},\ }\href {https://doi.org/10.1126/science.1139831} {\bibfield  {journal} {\bibinfo  {journal} {Science}\ }\textbf {\bibinfo {volume} {316}},\ \bibinfo {pages} {1312} (\bibinfo {year} {2007})},\ \Eprint {https://arxiv.org/abs/https://www.science.org/doi/pdf/10.1126/science.1139831} {https://www.science.org/doi/pdf/10.1126/science.1139831} \BibitemShut {NoStop}%
\bibitem [{\citenamefont {{Wu}}\ \emph {et~al.}(2020)\citenamefont {{Wu}}, \citenamefont {{Liang}}, \citenamefont {{Tian}}, \citenamefont {{Yang}}, \citenamefont {{Chen}}, \citenamefont {{Liu}}, \citenamefont {{Khoon Tey}},\ and\ \citenamefont {{You}}}]{2020arXiv201210614W}%
  \BibitemOpen
  \bibfield  {author} {\bibinfo {author} {\bibfnamefont {X.}~\bibnamefont {{Wu}}}, \bibinfo {author} {\bibfnamefont {X.}~\bibnamefont {{Liang}}}, \bibinfo {author} {\bibfnamefont {Y.}~\bibnamefont {{Tian}}}, \bibinfo {author} {\bibfnamefont {F.}~\bibnamefont {{Yang}}}, \bibinfo {author} {\bibfnamefont {C.}~\bibnamefont {{Chen}}}, \bibinfo {author} {\bibfnamefont {Y.-C.}\ \bibnamefont {{Liu}}}, \bibinfo {author} {\bibfnamefont {M.}~\bibnamefont {{Khoon Tey}}},\ and\ \bibinfo {author} {\bibfnamefont {L.}~\bibnamefont {{You}}},\ }\href@noop {} {\bibfield  {journal} {\bibinfo  {journal} {arXiv e-prints}\ ,\ \bibinfo {eid} {arXiv:2012.10614}} (\bibinfo {year} {2020})},\ \Eprint {https://arxiv.org/abs/2012.10614} {arXiv:2012.10614 [quant-ph]} \BibitemShut {NoStop}%
\bibitem [{\citenamefont {Kassal}\ \emph {et~al.}(2011)\citenamefont {Kassal}, \citenamefont {Whitfield}, \citenamefont {Perdomo-Ortiz}, \citenamefont {Yung},\ and\ \citenamefont {Aspuru-Guzik}}]{Kassal2011}%
  \BibitemOpen
  \bibfield  {author} {\bibinfo {author} {\bibfnamefont {I.}~\bibnamefont {Kassal}}, \bibinfo {author} {\bibfnamefont {J.~D.}\ \bibnamefont {Whitfield}}, \bibinfo {author} {\bibfnamefont {A.}~\bibnamefont {Perdomo-Ortiz}}, \bibinfo {author} {\bibfnamefont {M.-H.}\ \bibnamefont {Yung}},\ and\ \bibinfo {author} {\bibfnamefont {A.}~\bibnamefont {Aspuru-Guzik}},\ }\href {https://doi.org/10.1146/annurev-physchem-032210-103512} {\bibfield  {journal} {\bibinfo  {journal} {Annual Review of Physical Chemistry}\ }\textbf {\bibinfo {volume} {62}},\ \bibinfo {pages} {185} (\bibinfo {year} {2011})},\ \Eprint {https://arxiv.org/abs/https://doi.org/10.1146/annurev-physchem-032210-103512} {https://doi.org/10.1146/annurev-physchem-032210-103512} \BibitemShut {NoStop}%
\bibitem [{\citenamefont {McClean}\ \emph {et~al.}(2016)\citenamefont {McClean}, \citenamefont {Romero}, \citenamefont {Babbush},\ and\ \citenamefont {Aspuru-Guzik}}]{McClean_2016}%
  \BibitemOpen
  \bibfield  {author} {\bibinfo {author} {\bibfnamefont {J.~R.}\ \bibnamefont {McClean}}, \bibinfo {author} {\bibfnamefont {J.}~\bibnamefont {Romero}}, \bibinfo {author} {\bibfnamefont {R.}~\bibnamefont {Babbush}},\ and\ \bibinfo {author} {\bibfnamefont {A.}~\bibnamefont {Aspuru-Guzik}},\ }\href {https://doi.org/10.1088/1367-2630/18/2/023023} {\bibfield  {journal} {\bibinfo  {journal} {New Journal of Physics}\ }\textbf {\bibinfo {volume} {18}},\ \bibinfo {pages} {023023} (\bibinfo {year} {2016})}\BibitemShut {NoStop}%
\bibitem [{\citenamefont {{Grimsley}}\ \emph {et~al.}(2019)\citenamefont {{Grimsley}}, \citenamefont {{Economou}}, \citenamefont {{Barnes}},\ and\ \citenamefont {{Mayhall}}}]{2019NatCo..10.3007G}%
  \BibitemOpen
  \bibfield  {author} {\bibinfo {author} {\bibfnamefont {H.~R.}\ \bibnamefont {{Grimsley}}}, \bibinfo {author} {\bibfnamefont {S.~E.}\ \bibnamefont {{Economou}}}, \bibinfo {author} {\bibfnamefont {E.}~\bibnamefont {{Barnes}}},\ and\ \bibinfo {author} {\bibfnamefont {N.~J.}\ \bibnamefont {{Mayhall}}},\ }\href {https://doi.org/10.1038/s41467-019-10988-2} {\bibfield  {journal} {\bibinfo  {journal} {Nature Communications}\ }\textbf {\bibinfo {volume} {10}},\ \bibinfo {eid} {3007} (\bibinfo {year} {2019})},\ \Eprint {https://arxiv.org/abs/1812.11173} {arXiv:1812.11173 [quant-ph]} \BibitemShut {NoStop}%
\bibitem [{\citenamefont {{Parrish}}\ \emph {et~al.}(2019)\citenamefont {{Parrish}}, \citenamefont {{Hohenstein}}, \citenamefont {{McMahon}},\ and\ \citenamefont {{Martinez}}}]{2019arXiv190608728P}%
  \BibitemOpen
  \bibfield  {author} {\bibinfo {author} {\bibfnamefont {R.~M.}\ \bibnamefont {{Parrish}}}, \bibinfo {author} {\bibfnamefont {E.~G.}\ \bibnamefont {{Hohenstein}}}, \bibinfo {author} {\bibfnamefont {P.~L.}\ \bibnamefont {{McMahon}}},\ and\ \bibinfo {author} {\bibfnamefont {T.~J.}\ \bibnamefont {{Martinez}}},\ }\href@noop {} {\bibfield  {journal} {\bibinfo  {journal} {arXiv e-prints}\ ,\ \bibinfo {eid} {arXiv:1906.08728}} (\bibinfo {year} {2019})},\ \Eprint {https://arxiv.org/abs/1906.08728} {arXiv:1906.08728 [quant-ph]} \BibitemShut {NoStop}%
\bibitem [{\citenamefont {{Khoshaman}}\ \emph {et~al.}(2019)\citenamefont {{Khoshaman}}, \citenamefont {{Vinci}}, \citenamefont {{Denis}}, \citenamefont {{Andriyash}}, \citenamefont {{Sadeghi}},\ and\ \citenamefont {{Amin}}}]{2019QS&T....4a4001K}%
  \BibitemOpen
  \bibfield  {author} {\bibinfo {author} {\bibfnamefont {A.}~\bibnamefont {{Khoshaman}}}, \bibinfo {author} {\bibfnamefont {W.}~\bibnamefont {{Vinci}}}, \bibinfo {author} {\bibfnamefont {B.}~\bibnamefont {{Denis}}}, \bibinfo {author} {\bibfnamefont {E.}~\bibnamefont {{Andriyash}}}, \bibinfo {author} {\bibfnamefont {H.}~\bibnamefont {{Sadeghi}}},\ and\ \bibinfo {author} {\bibfnamefont {M.~H.}\ \bibnamefont {{Amin}}},\ }\href {https://doi.org/10.1088/2058-9565/aada1f} {\bibfield  {journal} {\bibinfo  {journal} {Quantum Science and Technology}\ }\textbf {\bibinfo {volume} {4}},\ \bibinfo {pages} {014001} (\bibinfo {year} {2019})},\ \Eprint {https://arxiv.org/abs/1802.05779} {arXiv:1802.05779 [quant-ph]} \BibitemShut {NoStop}%
\bibitem [{\citenamefont {{Havl{\'\i}{\v{c}}ek}}\ \emph {et~al.}(2019)\citenamefont {{Havl{\'\i}{\v{c}}ek}}, \citenamefont {{C{\'o}rcoles}}, \citenamefont {{Temme}}, \citenamefont {{Harrow}}, \citenamefont {{Kandala}}, \citenamefont {{Chow}},\ and\ \citenamefont {{Gambetta}}}]{2019Natur.567..209H}%
  \BibitemOpen
  \bibfield  {author} {\bibinfo {author} {\bibfnamefont {V.}~\bibnamefont {{Havl{\'\i}{\v{c}}ek}}}, \bibinfo {author} {\bibfnamefont {A.~D.}\ \bibnamefont {{C{\'o}rcoles}}}, \bibinfo {author} {\bibfnamefont {K.}~\bibnamefont {{Temme}}}, \bibinfo {author} {\bibfnamefont {A.~W.}\ \bibnamefont {{Harrow}}}, \bibinfo {author} {\bibfnamefont {A.}~\bibnamefont {{Kandala}}}, \bibinfo {author} {\bibfnamefont {J.~M.}\ \bibnamefont {{Chow}}},\ and\ \bibinfo {author} {\bibfnamefont {J.~M.}\ \bibnamefont {{Gambetta}}},\ }\href {https://doi.org/10.1038/s41586-019-0980-2} {\bibfield  {journal} {\bibinfo  {journal} {\nat}\ }\textbf {\bibinfo {volume} {567}},\ \bibinfo {pages} {209} (\bibinfo {year} {2019})},\ \Eprint {https://arxiv.org/abs/1804.11326} {arXiv:1804.11326 [quant-ph]} \BibitemShut {NoStop}%
\bibitem [{\citenamefont {{Benedetti}}\ \emph {et~al.}(2021)\citenamefont {{Benedetti}}, \citenamefont {{Coyle}}, \citenamefont {{Fiorentini}}, \citenamefont {{Lubasch}},\ and\ \citenamefont {{Rosenkranz}}}]{2021PhRvP..16d4057B}%
  \BibitemOpen
  \bibfield  {author} {\bibinfo {author} {\bibfnamefont {M.}~\bibnamefont {{Benedetti}}}, \bibinfo {author} {\bibfnamefont {B.}~\bibnamefont {{Coyle}}}, \bibinfo {author} {\bibfnamefont {M.}~\bibnamefont {{Fiorentini}}}, \bibinfo {author} {\bibfnamefont {M.}~\bibnamefont {{Lubasch}}},\ and\ \bibinfo {author} {\bibfnamefont {M.}~\bibnamefont {{Rosenkranz}}},\ }\href {https://doi.org/10.1103/PhysRevApplied.16.044057} {\bibfield  {journal} {\bibinfo  {journal} {Physical Review Applied}\ }\textbf {\bibinfo {volume} {16}},\ \bibinfo {eid} {044057} (\bibinfo {year} {2021})},\ \Eprint {https://arxiv.org/abs/2103.06720} {arXiv:2103.06720 [quant-ph]} \BibitemShut {NoStop}%
\bibitem [{\citenamefont {{Abel}}\ \emph {et~al.}(2022)\citenamefont {{Abel}}, \citenamefont {{Criado}},\ and\ \citenamefont {{Spannowsky}}}]{2022PhRvA.106b2601A}%
  \BibitemOpen
  \bibfield  {author} {\bibinfo {author} {\bibfnamefont {S.}~\bibnamefont {{Abel}}}, \bibinfo {author} {\bibfnamefont {J.~C.}\ \bibnamefont {{Criado}}},\ and\ \bibinfo {author} {\bibfnamefont {M.}~\bibnamefont {{Spannowsky}}},\ }\href {https://doi.org/10.1103/PhysRevA.106.022601} {\bibfield  {journal} {\bibinfo  {journal} {\pra}\ }\textbf {\bibinfo {volume} {106}},\ \bibinfo {eid} {022601} (\bibinfo {year} {2022})},\ \Eprint {https://arxiv.org/abs/2202.11727} {arXiv:2202.11727 [quant-ph]} \BibitemShut {NoStop}%
\bibitem [{\citenamefont {{Wang}}\ \emph {et~al.}(2020)\citenamefont {{Wang}}, \citenamefont {{Ashida}},\ and\ \citenamefont {{Ueda}}}]{2020PhRvL.125j0401W}%
  \BibitemOpen
  \bibfield  {author} {\bibinfo {author} {\bibfnamefont {Z.~T.}\ \bibnamefont {{Wang}}}, \bibinfo {author} {\bibfnamefont {Y.}~\bibnamefont {{Ashida}}},\ and\ \bibinfo {author} {\bibfnamefont {M.}~\bibnamefont {{Ueda}}},\ }\href {https://doi.org/10.1103/PhysRevLett.125.100401} {\bibfield  {journal} {\bibinfo  {journal} {\prl}\ }\textbf {\bibinfo {volume} {125}},\ \bibinfo {eid} {100401} (\bibinfo {year} {2020})},\ \Eprint {https://arxiv.org/abs/1910.09200} {arXiv:1910.09200 [quant-ph]} \BibitemShut {NoStop}%
\bibitem [{\citenamefont {{Kwak}}\ \emph {et~al.}(2022)\citenamefont {{Kwak}}, \citenamefont {{Yun}}, \citenamefont {{Pyoung Kim}}, \citenamefont {{Cho}}, \citenamefont {{Choi}}, \citenamefont {{Jung}},\ and\ \citenamefont {{Kim}}}]{2022arXiv220211200K}%
  \BibitemOpen
  \bibfield  {author} {\bibinfo {author} {\bibfnamefont {Y.}~\bibnamefont {{Kwak}}}, \bibinfo {author} {\bibfnamefont {W.~J.}\ \bibnamefont {{Yun}}}, \bibinfo {author} {\bibfnamefont {J.}~\bibnamefont {{Pyoung Kim}}}, \bibinfo {author} {\bibfnamefont {H.}~\bibnamefont {{Cho}}}, \bibinfo {author} {\bibfnamefont {M.}~\bibnamefont {{Choi}}}, \bibinfo {author} {\bibfnamefont {S.}~\bibnamefont {{Jung}}},\ and\ \bibinfo {author} {\bibfnamefont {J.}~\bibnamefont {{Kim}}},\ }\href@noop {} {\bibfield  {journal} {\bibinfo  {journal} {arXiv e-prints}\ ,\ \bibinfo {eid} {arXiv:2202.11200}} (\bibinfo {year} {2022})},\ \Eprint {https://arxiv.org/abs/2202.11200} {arXiv:2202.11200 [quant-ph]} \BibitemShut {NoStop}%
\bibitem [{\citenamefont {{Yang}}\ \emph {et~al.}(2022)\citenamefont {{Yang}}, \citenamefont {{Xiao}},\ and\ \citenamefont {{Yu}}}]{2022arXiv220608316Y}%
  \BibitemOpen
  \bibfield  {author} {\bibinfo {author} {\bibfnamefont {D.}~\bibnamefont {{Yang}}}, \bibinfo {author} {\bibfnamefont {Z.}~\bibnamefont {{Xiao}}},\ and\ \bibinfo {author} {\bibfnamefont {W.}~\bibnamefont {{Yu}}},\ }\href@noop {} {\bibfield  {journal} {\bibinfo  {journal} {arXiv e-prints}\ ,\ \bibinfo {eid} {arXiv:2206.08316}} (\bibinfo {year} {2022})},\ \Eprint {https://arxiv.org/abs/2206.08316} {arXiv:2206.08316 [cs.LG]} \BibitemShut {NoStop}%
\bibitem [{\citenamefont {Mitarai}\ \emph {et~al.}(2018)\citenamefont {Mitarai}, \citenamefont {Negoro}, \citenamefont {Kitagawa},\ and\ \citenamefont {Fujii}}]{PhysRevA.98.032309}%
  \BibitemOpen
  \bibfield  {author} {\bibinfo {author} {\bibfnamefont {K.}~\bibnamefont {Mitarai}}, \bibinfo {author} {\bibfnamefont {M.}~\bibnamefont {Negoro}}, \bibinfo {author} {\bibfnamefont {M.}~\bibnamefont {Kitagawa}},\ and\ \bibinfo {author} {\bibfnamefont {K.}~\bibnamefont {Fujii}},\ }\href {https://doi.org/10.1103/PhysRevA.98.032309} {\bibfield  {journal} {\bibinfo  {journal} {Phys. Rev. A}\ }\textbf {\bibinfo {volume} {98}},\ \bibinfo {pages} {032309} (\bibinfo {year} {2018})}\BibitemShut {NoStop}%
\bibitem [{\citenamefont {{Liu}}\ and\ \citenamefont {{Wang}}(2018)}]{2018arXiv180404168L}%
  \BibitemOpen
  \bibfield  {author} {\bibinfo {author} {\bibfnamefont {J.-G.}\ \bibnamefont {{Liu}}}\ and\ \bibinfo {author} {\bibfnamefont {L.}~\bibnamefont {{Wang}}},\ }\href@noop {} {\bibfield  {journal} {\bibinfo  {journal} {arXiv e-prints}\ ,\ \bibinfo {eid} {arXiv:1804.04168}} (\bibinfo {year} {2018})},\ \Eprint {https://arxiv.org/abs/1804.04168} {arXiv:1804.04168 [quant-ph]} \BibitemShut {NoStop}%
\bibitem [{\citenamefont {Coyle}\ \emph {et~al.}(2020)\citenamefont {Coyle}, \citenamefont {Mills}, \citenamefont {Danos},\ and\ \citenamefont {Kashefi}}]{coyle_born_2020}%
  \BibitemOpen
  \bibfield  {author} {\bibinfo {author} {\bibfnamefont {B.}~\bibnamefont {Coyle}}, \bibinfo {author} {\bibfnamefont {D.}~\bibnamefont {Mills}}, \bibinfo {author} {\bibfnamefont {V.}~\bibnamefont {Danos}},\ and\ \bibinfo {author} {\bibfnamefont {E.}~\bibnamefont {Kashefi}},\ }\href {https://doi.org/10.1038/s41534-020-00288-9} {\bibfield  {journal} {\bibinfo  {journal} {npj Quantum Information}\ }\textbf {\bibinfo {volume} {6}},\ \bibinfo {pages} {60} (\bibinfo {year} {2020})}\BibitemShut {NoStop}%
\bibitem [{\citenamefont {{Nakanishi}}\ \emph {et~al.}(2018)\citenamefont {{Nakanishi}}, \citenamefont {{Mitarai}},\ and\ \citenamefont {{Fujii}}}]{2018arXiv181009434N}%
  \BibitemOpen
  \bibfield  {author} {\bibinfo {author} {\bibfnamefont {K.~M.}\ \bibnamefont {{Nakanishi}}}, \bibinfo {author} {\bibfnamefont {K.}~\bibnamefont {{Mitarai}}},\ and\ \bibinfo {author} {\bibfnamefont {K.}~\bibnamefont {{Fujii}}},\ }\href@noop {} {\bibfield  {journal} {\bibinfo  {journal} {arXiv e-prints}\ ,\ \bibinfo {eid} {arXiv:1810.09434}} (\bibinfo {year} {2018})},\ \Eprint {https://arxiv.org/abs/1810.09434} {arXiv:1810.09434 [quant-ph]} \BibitemShut {NoStop}%
\bibitem [{\citenamefont {{Romero}}\ \emph {et~al.}(2017)\citenamefont {{Romero}}, \citenamefont {{Babbush}}, \citenamefont {{McClean}}, \citenamefont {{Hempel}}, \citenamefont {{Love}},\ and\ \citenamefont {{Aspuru-Guzik}}}]{2017arXiv170102691R}%
  \BibitemOpen
  \bibfield  {author} {\bibinfo {author} {\bibfnamefont {J.}~\bibnamefont {{Romero}}}, \bibinfo {author} {\bibfnamefont {R.}~\bibnamefont {{Babbush}}}, \bibinfo {author} {\bibfnamefont {J.~R.}\ \bibnamefont {{McClean}}}, \bibinfo {author} {\bibfnamefont {C.}~\bibnamefont {{Hempel}}}, \bibinfo {author} {\bibfnamefont {P.}~\bibnamefont {{Love}}},\ and\ \bibinfo {author} {\bibfnamefont {A.}~\bibnamefont {{Aspuru-Guzik}}},\ }\href@noop {} {\bibfield  {journal} {\bibinfo  {journal} {arXiv e-prints}\ ,\ \bibinfo {eid} {arXiv:1701.02691}} (\bibinfo {year} {2017})},\ \Eprint {https://arxiv.org/abs/1701.02691} {arXiv:1701.02691 [quant-ph]} \BibitemShut {NoStop}%
\bibitem [{\citenamefont {Fletcher}(2013)}]{doi:10.1002/9781118723203.ch3}%
  \BibitemOpen
  \bibfield  {author} {\bibinfo {author} {\bibfnamefont {R.}~\bibnamefont {Fletcher}},\ }\bibinfo {title} {Newton-like methods},\ in\ \href {https://doi.org/10.1002/9781118723203.ch3} {\emph {\bibinfo {booktitle} {Practical Methods of Optimization}}}\ (\bibinfo  {publisher} {John Wiley \& Sons, Ltd},\ \bibinfo {year} {2013})\ Chap.~\bibinfo {chapter} {3}, pp.\ \bibinfo {pages} {44--79},\ \Eprint {https://arxiv.org/abs/https://onlinelibrary.wiley.com/doi/pdf/10.1002/9781118723203.ch3} {https://onlinelibrary.wiley.com/doi/pdf/10.1002/9781118723203.ch3} \BibitemShut {NoStop}%
\bibitem [{\citenamefont {Benedetti}\ \emph {et~al.}(2019)\citenamefont {Benedetti}, \citenamefont {Garcia-Pintos}, \citenamefont {Perdomo}, \citenamefont {Leyton-Ortega}, \citenamefont {Nam},\ and\ \citenamefont {Perdomo-Ortiz}}]{benedetti_generative_2019}%
  \BibitemOpen
  \bibfield  {author} {\bibinfo {author} {\bibfnamefont {M.}~\bibnamefont {Benedetti}}, \bibinfo {author} {\bibfnamefont {D.}~\bibnamefont {Garcia-Pintos}}, \bibinfo {author} {\bibfnamefont {O.}~\bibnamefont {Perdomo}}, \bibinfo {author} {\bibfnamefont {V.}~\bibnamefont {Leyton-Ortega}}, \bibinfo {author} {\bibfnamefont {Y.}~\bibnamefont {Nam}},\ and\ \bibinfo {author} {\bibfnamefont {A.}~\bibnamefont {Perdomo-Ortiz}},\ }\href {https://doi.org/10.1038/s41534-019-0157-8} {\bibfield  {journal} {\bibinfo  {journal} {npj Quantum Information}\ }\textbf {\bibinfo {volume} {5}},\ \bibinfo {pages} {45} (\bibinfo {year} {2019})}\BibitemShut {NoStop}%
\bibitem [{\citenamefont {{Shtanko}}\ and\ \citenamefont {{Movassagh}}(2021)}]{2021arXiv211214688S}%
  \BibitemOpen
  \bibfield  {author} {\bibinfo {author} {\bibfnamefont {O.}~\bibnamefont {{Shtanko}}}\ and\ \bibinfo {author} {\bibfnamefont {R.}~\bibnamefont {{Movassagh}}},\ }\href@noop {} {\bibfield  {journal} {\bibinfo  {journal} {arXiv e-prints}\ ,\ \bibinfo {eid} {arXiv:2112.14688}} (\bibinfo {year} {2021})},\ \Eprint {https://arxiv.org/abs/2112.14688} {arXiv:2112.14688 [quant-ph]} \BibitemShut {NoStop}%
\bibitem [{\citenamefont {{Han}}\ \emph {et~al.}(2021)\citenamefont {{Han}}, \citenamefont {{Cai}}, \citenamefont {{Hu}}, \citenamefont {{Mu}}, \citenamefont {{Ma}}, \citenamefont {{Xu}}, \citenamefont {{Wang}}, \citenamefont {{Wang}}, \citenamefont {{Song}}, \citenamefont {{Zou}},\ and\ \citenamefont {{Sun}}}]{2021PhRvL.127b0504H}%
  \BibitemOpen
  \bibfield  {author} {\bibinfo {author} {\bibfnamefont {J.}~\bibnamefont {{Han}}}, \bibinfo {author} {\bibfnamefont {W.}~\bibnamefont {{Cai}}}, \bibinfo {author} {\bibfnamefont {L.}~\bibnamefont {{Hu}}}, \bibinfo {author} {\bibfnamefont {X.}~\bibnamefont {{Mu}}}, \bibinfo {author} {\bibfnamefont {Y.}~\bibnamefont {{Ma}}}, \bibinfo {author} {\bibfnamefont {Y.}~\bibnamefont {{Xu}}}, \bibinfo {author} {\bibfnamefont {W.}~\bibnamefont {{Wang}}}, \bibinfo {author} {\bibfnamefont {H.}~\bibnamefont {{Wang}}}, \bibinfo {author} {\bibfnamefont {Y.~P.}\ \bibnamefont {{Song}}}, \bibinfo {author} {\bibfnamefont {C.~L.}\ \bibnamefont {{Zou}}},\ and\ \bibinfo {author} {\bibfnamefont {L.}~\bibnamefont {{Sun}}},\ }\href {https://doi.org/10.1103/PhysRevLett.127.020504} {\bibfield  {journal} {\bibinfo  {journal} {\prl}\ }\textbf {\bibinfo {volume} {127}},\ \bibinfo {eid} {020504} (\bibinfo {year} {2021})},\ \Eprint {https://arxiv.org/abs/2108.02395} {arXiv:2108.02395 [quant-ph]} \BibitemShut {NoStop}%
\end{thebibliography}%

\end{document}